\documentclass[twocolumn]{aastex631}

\shorttitle{Recurrent Nova T Pyxidis is Not a Supernova Progenitor}
\shortauthors{Schaefer}
\graphicspath{{./}{figures/}}

\begin{document}
\title{Orbital Period Changes of Recurrent Nova T Pyxidis Demonstrate that M$_{\rm ejecta}$$\gg$11.3$\times$M$_{\rm accreted}$ and Is Not a Type Ia Supernova Progenitor}

\author[0000-0002-2659-8763]{Bradley E. Schaefer}
\affiliation{Department of Physics and Astronomy,
Louisiana State University, Baton Rouge, LA 70803, USA}

\begin{abstract}

Recurrent nova (RN) T Pyxidis (T Pyx) has a complex history of mass accreting-onto and ejection-from the white dwarf, with a classical nova eruption around 1866 kick-starting a RN-phase with six RN eruptions from 1890--2011.  T Pyx is a primary progenitor candidate for Type Ia supernovae (SNIa).  This is chiefly a question of whether the mass accreted by the white dwarf ($M_{\rm accreted}$) is more-or-less than the mass ejected by the nova eruptions ($M_{\rm ejecta}$) over the entire eruption cycle.  Prior attempts to measure $M_{\rm ejecta}$ from the traditional methods have a scatter of $>$130$\times$, so only a new technique can provide a measure of adequate accuracy and reliability.  This new technique is the timing experiment  of measuring the orbital period from 1986 to 2025, where the period increased by $+$50.3$\pm$7.9 parts-per-million across the 2011 eruption.  With simple and sure physics, the best estimate for the mass ejected by one RN event is $>$2400$\times$10$^{-7}$ M$_{\odot}$, with an extreme inviolate limit of $\gg$354$\times$10$^{-7}$ M$_{\odot}$.  Over all eruptions in a cycle, $M_{ejecta}$$>$17120$\times$10$^{-7}$ M$_{\odot}$, with an inviolate limit of $M_{ejecta}$$\gg$2144$\times$10$^{-7}$ M$_{\odot}$.  Over the full eruption cycle, the white dwarf accreted 220$\times$10$^{-7}$ M$_{\odot}$.  So M$_{\rm ejecta}$$\gg$11.3$\times$M$_{\rm accreted}$, and T Pyx can never become a SNIa.  This paper is the seventh in a series proving that each of various popular candidate SNIa progenitors cannot possibly evolve to a supernova; including V445 Pup, U Sco, T CrB, all symbiotic stars, FQ Cir, V1405 Cas, and now T Pyx.

\end{abstract}

\section{INTRODUCTION}

In January of 2010, the famous recurrent nova (RN) T Pyx made widespread lurid international news from a conference talk at an {\it American Astronomical Society} meeting and its press release (Sion, Godon, and McClain 2010).  The talk title was ``The Long Overdue Recurrent Nova T Pyxidis: Soon to be Type Ia Supernova?'', and the press release concluded ``gamma radiation emitted by the supernova would fry the Earth, dumping as much gamma radiation ($\sim$100,000 erg/square centimeter) into our planet, which is equivalent to the gamma ray input of 1000 solar flares simultaneously.''  This particular astrophysics is broadly rejected as erroneous on multiple grounds, and the claim never made it into a journal article.  Nevertheless, this highlights the important question of whether T Pyx can or will become a Type Ia supernova (SNIa).

The nature of the progenitor for SNIa has been one of the grand challenge problems of high importance throughout astrophysics (Livio 2000, Maoz, Mannucci, \& Nelemans 2014, Ruiter \& Seitenzahl 2025).  SNIa are the group of `standardizable candles' used to discover Dark Energy.  The progenitors of these explosions are certainly the thermonuclear burning of carbon/oxygen (CO) white dwarfs (WDs).  Before 1984\footnote{In 1984, Webbink (1984) and Iben \& Tutukov (1984) proposed the idea that SNIa could be caused by the in-spiral of a close binary consisting of {\it two} CO WDs.  This is now called the Double-Degenerate model.  These 1984 papers laid out most of the debate between the competing models, Single-Degenerate versus Double-Degenerate.  Since 1984, the debate has been waged over a wide range of evidences, going many levels deep, often with vehement discussion amongst advocates with set resolve.  T Pyx has entered this debate as one of the exemplars for the Single-Degenerate advocates. }, the only idea on the progenitor nature of what we now call Type Ia supernovae was that a close interacting binary with {\it one} WD accreted gas from its companion star, until the WD neared the Chandrasekhar mass, whereupon its collapse ignites runaway nuclear burning of the CO, releasing the energy that blows apart the WD as a supernova (Trimble 1984).  This is the so-called Single-Degenerate (SD) model.  The SD progenitors are cataclysmic variables (CVs), from which the RNe are the best candidates for progenitors, because they necessarily have the highest-mass WDs and have near-maximal accretion onto the WDs.  As one of the famous RNe, T Pyx became one of the best candidates to solve the all-important SNIa Progenitor Problem.

The fate of T Pyx as a Type Ia supernova has been long debated.  Here is a selection of papers offering conclusions:  Della Valle \& Livio (1996) {\it presumed} that the RN WDs are increasing in mass, then made a crude estimate of their birth rate, finding this to be a few percent of the SNIa rate.  Knigge, King, \& Patterson (2000) concluded that T Pyx is a wind-driven supersoft X-ray source that will commit `assisted stellar suicide' as a Type Ia supernova, given that the total system mass is larger than the Chandrasekhar mass.  Selvelli et al. (2008) title their broad-ranging analysis as ``The secrets of T Pyxidis  II. A recurrent nova that will not become a SN Ia''.

So the critical question is whether T Pyx can become a SNIa?  This question comes down to whether the WD mass ($M_{\rm WD}$) is increasing over evolutionary timescales.  Simplistically, the high accretion rate onto the WD (required for the short RN recurrence timescale) shows that the WD is increasing mass fast.  But the WD is also losing mass each nova eruption.  So the critical question becomes one of balancing the mass accreted over each eruption cycle ($M_{\rm accreted}$) against the mass ejected in that eruption cycle ($M_{\rm ejecta}$).  If $M_{\rm accreted}$$<$$M_{\rm ejecta}$, then the T Pyx WD is not approaching the Chandrasekhar mass and it cannot become a SNIa\footnote{A second requirement for T Pyx to become a SNIa is that its WD must be of CO composition.  Large amounts of carbon are needed as the energy source for the normal SNIa explosion.  The alternative is a WD with predominantly oxygen/neon (ONe) composition.  ONe WDs are exclusively formed only with masses $\gtrsim$1.2 $M_{\odot}$ or so.  If a WD has an ONe composition, then available nuclear energy upon collapse is too small to power normal SNIa.  In this case, the WD would suffer a quiet `accretion induced collapse' to make a neutron star.  The presence of an underlying ONe WD is primarily recognizable when the nova dredges-up neon-rich material from the WD core and mantle, with this being seen as a neon nova, with a high abundance of neon in the ejecta.  T Pyx is not a neon nova.  Nevertheless, the T Pyx WD could still be an ONe WD if the RN system is fairly young.  In this case, the nova would be eroding away the outer mantle of the WD, and ONe WDs have the outer regions primarily of carbon and oxygen composition (de Ger\'{o}nimo et al. 2019).  For T Pyx in particular, Chomiuk et al. (2014) prove a super-solar abundance of oxygen in the ejecta, with the requirement that the ejecta must include large amounts of dredged-up WD material.  Only after many nova events dredge-up and eject all of the WD mantle material will the high neon abundance material of the WD interior be revealed as a neon nova.  If $M_{\rm accreted}$$<$$M_{\rm ejecta}$ for T Pyx, then the WD is losing mass over time, and it must have formed as a near-Chandrasekhar-mass WD, which necessarily has an ONe composition.  So if $M_{\rm accreted}$$<$$M_{\rm ejecta}$, then we have two strong proofs that T Pyx cannot become a SNIa, first because the WD mass is {\it not} increasing to the Chandrasekhar mass, and second because the WD has an ONe composition and cannot explode as a normal SNIa.  }.  In practice, $M_{\rm accreted}$ can be measured to usable accuracy.  In practice, all the old and traditional methods to measure $M_{\rm ejecta}$ have real uncertainties 2--3 orders of magnitude in size, see Section 3.1.  For T Pyx in particular, published measures for $M_{\rm ejecta}$ range by a factor of over 130$\times$, so the old methods have failed.  So the critical question comes down to finding some new, reliable, and accurate method to measure $M_{\rm ejecta}$.

In this paper, I will measure both $M_{\rm accreted}$ and $M_{\rm ejecta}$, to resolve the question of whether T Pyx can possibly become a Type Ia supernova.  For this, I will measure the ejected mass by means of my measure of the orbital period change ($\Delta P$) across the 2011 RN eruption.  This method is a simple timing experiment with simple and reliable physics, where the inputs are accurately known\footnote{Importantly, the $\Delta P$ method has no dependency on distance, extinction, filling factors, temperatures, or models of complex situations.  The physics used is simple, with no loopholes, see the full derivation in Equations 2--5, where the only input is from Kepler's Law and the conservation of angular momentum, all with no complications.  There is no way around the strict limit derived in Equation 5.}.  This method only works for novae that have positive measured $\Delta P$, and that only with many decades of a massive and relentless campaign of photometric time series.  T Pyx is one of the few novae for which this method is possible.  Both Patterson, Oksanen, \& Kemp (2017) and Schaefer (2023b) have measured the $\Delta P$ for the 2011 eruption, and this serves as basic input for the analysis below.  The situation for T Pyx is complex because the eruption cycles includes many RN eruptions as triggered by a classical nova eruption, while the accretion rate changes by 4 orders-of-magnitude from the long faint quiescence before the classical nova trigger to the maximal accretion rate near the start of the RN episodes and the unique decline in accretion over the two-century RN episodes.  (All prior analyses have just considered the {\it one} eruption and less than half of the {\it one} preceding interval of quiescence, and such is only a small and un-representative fraction of the whole case.)  The task here is to sum up all the accretion from the 1866 classical nova eruption, through the entire RN-phase, and through the quiescent phase, and to sum up the ejected masses from all the RN and classical nova eruptions.  Only then can we test whether $M_{\rm accreted}$$<$$M_{\rm ejecta}$.  My T Pyx campaign started back in 1987 with the explicit purpose of testing whether T Pyx can become a SNIa (Schaefer et al. 1992), and this paper represents the culmination of this 38-year program.

This paper is the seventh of a series in the {\it Astrophysical Journal}, all of which prove that specific popular SNIa progenitor candidates cannot possibly evolve to become normal SNIa:  {\bf 1.~~}Schaefer (2025a) proves that the helium nova V445 Pup cannot become a supernova because its WD is {\it decreasing} in mass, as measured from the $\Delta P$ with $M_{\rm ejecta}$$\gg$0.001 $M_{\odot}$.  As the only known helium nova, V445 Pup has been the prototype for a whole class of supernova progenitor models.  {\bf 2.~~}Schaefer (2025c) proves that recurrent nova U Sco cannot become a SNIa because the measured $\Delta P$ has $M_{\rm ejecta}$=26$\times$$M_{\rm accreted}$.  U Sco is an exemplar of models taking RNe as the primary progenitors of SNIa.  {\bf 3.~~}Schaefer (2025d) proves that T CrB cannot possibly be a SNIa progenitor because the $\Delta P$ shows $M_{\rm ejecta}$=540$\times$$M_{\rm accreted}$.  T CrB is one of the premier and most-popular SNIa progenitor candidates because it is both a symbiotic star and a recurrent nova.  {\bf 4.~~}Schaefer (2025f) proves that symbiotic stars cannot provide any measurable fraction ($<$0.54\%) of normal SNIa, because zero out of 189 normal SNIa have any possibility of having a red giant companion star.  Symbiotic stars are another popular set of progenitor candidates, but they are completely ruled out.  {\bf 5.~~}Schaefer et al. (2025) proves that FQ Cir cannot be a supernova progenitor because the mass budget forces the WD to have ONe composition.   FQ Cir appeared as an ordinary classical nova eruption in 2021, but it turns out that the companion star is a 13.0 $M_{\odot}$ B0e star whose decretion disk is Roche lobe overflowing onto the accretion disk around the WD.  The only other known high-mass CV is a rapid-recurrent-nova in the Andromeda Galaxy, a class identified as ``the leading pre-explosion supernova Type Ia candidate system'' (Darnley et al. 2017).  {\bf 6.~~}Schaefer (2025g) proves that ordinary nova V1405 Cas has $M_{\rm ejecta}$$>$$M_{\rm accreted}$, again from my measured $\Delta P$.  Further, V1405 Cas is a neon nova and thus cannot be a SNIa progenitor.  V1405 Cas is one of the slowest novae, with one of the smallest $M_{\rm WD}$, and can serve as a test for progenitor candidates of low-mass CVs.  {\bf 7.~~}This paper proves that T Pyx cannot be a SNIa progenitor, because my $\Delta P$ measure shows $M_{\rm ejecta}$$\gg$11.3$\times$$M_{\rm accreted}$.  This is a greatly more complex result than previously appreciated, because account has to be made of the complex history before and after the classical nova event of 1866 and the subsequent century-long RN series.

\section{T PYX PROPERTIES}

\subsection{Eruptions}

T Pyx has known nova eruptions in the years 1890, 1902, 1920, 1944, 1967, and 2011.  The light curves of all these events are identical to within the usual measurement errors and small bumps (Schaefer 2010, Schaefer et al. 2013), while the spectral developments are also identical.  The short recurrence times of 12, 18, 24, 23, and 44 years are what define T Pyx as a recurrent nova.  These eruptions are the usual physical situations that are characteristic of recurrent nova events, with extremely fast expansion velocity of 5350 km/s, white dwarf mass of 1.33 $M_{\odot}$, light curve class of `P', and extremely high accretion up to 2$\times$10$^{-7}$ M$_{\odot}$/year in quiescence.

T Pyx is also surrounded by a unique nova shell that has been broken up into a myriad of knots, each fading and brightening (Shara et al. 1989, 1997, 2015, Schaefer, Pagnotta, \& Shara 2010).  Importantly, this shell is seen to be expanding slowly, greatly slower than any RN eruption.  The observed shell and knots cannot have come from the ejecta of any of the known RN eruptions.  A detailed comparison of the knot positions in the 1994/5 and 2007 images from {\it HST} show the nova shell to be expanding at 500--715 km/s (for a distance of 3500 pc).  Extrapolating back in time, the knots were ejected in the year 1866$\pm$5.  The total mass of the shell was estimated as $\sim$10$^{-4.5}$ $M_{\odot}$.  This high-mass and the slow-velocity prove that the ejection could not have come from any sort of a recurrent nova eruption.  Rather, the high-mass and slow-velocity are uniquely those of an ordinary classical nova eruption.  So the visible nova shell was ejected by T Pyx around the year 1866 with a velocity around 650 km/s.  Subsequent recurrent nova eruptions (with their extremely fast shells ejected) blow through the old-and-slow shell from c.1866, forming the observed knots from the expected Rayleigh-Taylor instabilities (Toraskar et al. 2013).  The later fast shells pass through the shredded c.1866 shell, shock ionizing the knots which brighten and then fade.

The confident scenario is that T Pyx had an ordinary classical nova eruption\footnote{The peak $V$-band absolute magnitude of all novae is $-$7.45 (with an RMS scatter of 1.33 mag, Schaefer 2022b), so along with a distance of 3599 pc and $E(B-V)$ of 0.25 (see below), the expected peak brightness of the 1866 ordinary nova eruption should be somewhere around $V$=6.1.  Unfortunately, back around 1866, there were no systematic observers of the sky in the southern hemisphere.  As for laypersons or skywatchers not looking for novae, the effective limit for catching an unexpected nova is around $V$=2, so there is no chance that the 1866 eruption would have been accidentally discovered.  Even back in the heyday of the Harvard plate archives, the probability for an undirected discovery of T Pyx was 20\% (Schaefer 2010), but there are no useful astronomical photographs before 1890.} around 1866, and this somehow initiated a series of fast recurrent nova eruptions from at least 1890 to 2011.  This required case suddenly becomes the dominant fact in all consideration of the evolution and fate of T Pyx.

\subsection{Distance}

The distance, $D$, is one of the most important property of a CV, because the derived luminosities and energies have strong dependencies.  For T Pyx in particular, for times before the {\it Gaia} parallax became available, many papers reported a distance measure for T Pyx, ranging from 1000 to 5000 pc.  Unfortunately, as given in fine detailed evaluations of each claim, Schaefer et al. (2013) demonstrates that ``all prior distance estimates to T Pyx are either erroneous, not applicable, and/or uselessly poor.''  For example, a popular but utterly-failed method is the so-called ``maximum magnitude--rate of decline'' (MMRD) relation, which proved to be bad for most nova populations (e.g., Kasliwal et al. 2011), proved by the {\it Gaia} distances to have a real scatter of 5 mags in the distance modulus (Schaefer 2022b), and proved to be merely a selection effect (Shafter, Clark, \& Hornoch 2023).  Another commonly-used method is to relate observed equivalent widths of interstellar lines to $D$, but such fails because the variation from star-to-star is always huge, even for the same distance, so the real uncertainties in calibration must be near one order-of-magnitude in distance.  Sion, Godon, \& McClain (2010) claim a distance of 1000 pc as ultimately based on the flux in the {\it HST} ultraviolet spectrum, but this estimate has large systematic problems because their disk models were made for WD masses $>$0.1 $M_{\odot}$ too small and accretion rates 50$\times$ too small, while the spectral energy distribution already proved that the ultraviolet emission is not dominantly from any disk light as assumed.  In the end, as detailed claim-by-claim in Schaefer et al. (2013), there are no correct claims for the T Pyx distance accurate to better than a factor of 3$\times$.

Fortunately, the {\it Gaia} satellite measured a reliable and accurate parallax for T Pyx.  To convert the parallax into $D$, the correct and required computation is a Bayesian calculation involving a prior for the space distribution of novae in the galactic disk population\footnote{{\it For nova distances}, an easy mistake is to simply use the distance quoted by the {\it Gaia} archive (distance\_gspphot), which is 4113.787$^{+40}_{-25}$ pc for T Pyx.  The problem is that the catalog assumes a distance scale of $L$$\approx$1000 pc for the EDSD distance prior, as based on their chemo-dynamical model of the Milky Way, as appropriate for random field stars (Bailer-Jones et al. 2018).  But T Pyx is not a random field star, nor is the galactic distribution of disk novae similar in scale as random field stars.  The {\it Gaia} catalog presumption pushed the T Pyx distance to a larger value than is best.}.  Schaefer 2022b used all the {\it Gaia} nova parallaxes to derive this space distribution to be an exponential in distance from the midplane, with a scale height of 140 pc.  Further, the priors should include previous measures of the T Pyx distance, and those error bars must be the realistic total error bars\footnote{The usually published error bars are only from the propagation of the input measurement uncertainties.  Rather the total error bars for use in the Bayesian priors, including systematic problems, are evaluated as in Schaefer (2018).}.  Schaefer (2022b) has performed the required calculations exhaustively for all 402 known galactic novae, with these distances including and superseding all prior distance estimates.  For T Pyx, the $D$ is 3599$^{+332}_{-230}$ parsecs.

\subsection{Extinction}

Bruch  \& Engel (1994) report $E(B-V)$=0.20.  Weight et al. (1994) quote a reddening of 0.08 mag from their study of novae in the near infrared. Shore et al. (2011) report $E(B-V)$$\approx$0.5$\pm$0.1, as based on the strengths of some diffuse interstellar bands, but Schaefer et al. (2013) use their given data to revise their measure to 0.47$\pm$0.23.  Schlafly \& Finkbeiner (2011) report an accurate and reliable total reddening of $E(B-V)$=0.235$\pm$0.009 for the entire column of gas and dust through the entire Milky Way.  With T Pyx being 610 pc above the galactic plane, far above nearly all the dust, the extinction to T Pyx can only be close to 0.235 mag.  Gilmozzi \& Selvelli (2007) used archival {\it IUE} spectra to deredden the usual 2175~\AA~ bump to measure $E(B-V)$=0.25$\pm$0.02, with this being the best method and the best measure of the extinction.  Schaefer et al. (2013) confirm the {\it IUE} value with the substantially poorer {\it Galex} spectrum.  In the end, here I adopt the Selvelli \& Gilmozzi value of $E(B-V)$=0.25$\pm$0.02.

\subsection{Spectral Energy Distribution}

A top quality spectral energy distribution (SED) can be constructed from the far ultraviolet to the mid-infrared.  Gilmozzi \& Selvelli (2007) used archival {\it IUE} spectra from 1260--3200~\AA~to fit the ultraviolet continuum to a power law with a best-fit $f_{\nu}$$\propto$$\nu^{0.33}$, with this being constant from 1980 to 1996.  Further, they extend their UV spectrum out with $BVRJ$ photometry, to get a best fit as $f_{\nu}$$\propto$$\nu^{0.9}$.  In the ultraviolet alone from the {\it Galex} spectrum of December 2005, Schaefer et al. (2013) found a good power law as $f_{\nu}$$\propto$$\nu^{0.25\pm0.03}$.  Further, with $UBVRIJHK$ photometry, Schaefer et al. (2013) find a good single power law SED that is close to $f_{\nu}$$\propto$$\nu^{1.0}$.  (The ultraviolet slope is just a small wiggle on the SED.)    So these independent measures from 2005 and 1980--1996 show a constant and consistent SED that is close to a $\nu^{1.0}$ power law, with no significant breaks anywhere.

The SED can tell about the dominant emission sources.  The observed SED ($f_{\nu}$$\propto$$\nu^{1.0}$) is greatly different from any blackbody, for which the Rayleigh-Jeans part is $f_{\nu}$$\propto$$\nu^{2.0}$.  So the SED has no significant contribution from the companion star, nor from any hot thermal source.  The SED for all accretion disks is $f_{\nu}$$\propto$$\nu^{1/3}$ near its peak, and rolling over to a Rayleigh-Jeans spectrum on the red side of the peak.  At no wavelength does the T Pyx SED have any non-negligible contribution from the accretion disk.  This demonstrates that the accretion disk is not contributing significantly to the luminosity in quiescence.  Indeed, Schaefer et al. (2013) proves that the observed SED cannot be made from the summation of any number of blackbodies.  Ordinary emission from the WD cannot provide this luminosity, while even a brightly-irradiated companion star is too cool and small to contribute significantly to the SED.  This means that there must be some additional source of high luminosity.  The power-law shape proves that the dominant emission mechanism must be non-thermal.  The only realistic possibility is from nuclear burning on the white dwarf, possibly providing a puffed-up hot envelope.  

Such a case was independently proposed by Patterson et al. (1998) and Knigge, King, \& Patterson (2000) in a good attempt to explain the near-maximal accretion rate on a CV at the bottom of the Period Gap.  They are making T Pyx to be a supersoft X-ray source, which drives a strong radiation-induced wind from the surface of the companion star.  The wind then feeds the high accretion rate that powers the nuclear burning, all in a feedback loop.  This feedback is weakening over the decades, leading to the fading in quiescence from $B$=13.95 in 1890 to $B$=16.06 in 2025.

\subsection{White Dwarf Mass}

Uthas, Knigge, \& Steeghs (2010) measure a radial velocity (RV) curve for T Pyx, and they report a WD mass of 0.7$\pm$0.2 M$_{\odot}$.  This is taken from RV measures of just one emission line (the double-peaked He II line at 4686~\AA), with this producing a nice RV plot where the binned RV curve has a scatter that is $\sim$15\% of the full amplitude.  They derive a semi-amplitude of $K$$\ge$17.9$\pm$1.6 km/s, where the measure is stated to be only a limit.  They state that their measure ``is subject to considerably systematic uncertainties''.  These are demonstrated by their systemic $\gamma$ velocities being time variable, by the K-velocities differing from line-to-line, and by their fit parameters not converging in their diagnostic diagrams.  Then, using a non-standard analysis, they can only make an estimate of the mass ratio ($q$=$M_{\rm comp}$/$M_{\rm WD}$) as 0.20$\pm$0.03.  Actually, they state that this is only a limit on $q$.  No orbital inclination is known, but such is apparently small.  With no measure of $M_{\rm WD}$ or $M_{\rm comp}$ specifically, they then adopted $M_{\rm comp}$=0.14$\pm$0.03 $M_{\odot}$ (from the stellar mass-radius relation) to derive their final 0.7$\pm$0.2 $M_{\odot}$.  This quoted error bar only covers the propagation from the uncertainties in $q$ and $M_{\rm comp}$.  Rather, the real total error bar must have a large contribution from the various systematic errors and assumptions, which apparently are huge.  And the quoted mass is only a lower limit.  As such, I conclude that this mass measure has no usable reliability\footnote{Popularly, RV mass measures are taken as the `gold standard'.  However, for novae and CVs, RV WD masses are notoriously poor.  Schaefer (2025e, Sections 2.1.1 and 2.1.5) details the deep and ubiquitous problems and counterexamples.  Part of the problem is that the emission lines are from the disk and do not map out the WD velocity, often showing large phase lags with respect to minimum times, while irradiation effects also bollux any accurate measure.  Schaefer (2025e) collected all 17 published RV WD masses for RNe, with this having the advantage that we know the masses for the RNe with recurrence time scales faster than 30 years must be 1.20--1.40 $M_{\odot}$ with no possibility of an exception.  The conclusion was ``This is an abysmal record for RV masses as a method. Of the 17 published RV masses for RNe, 4 are larger than the Chandrasekhar mass, 6 are impossibly small at $\leq$1.1 $M_{\odot}$, and three were hopelessly confused despite apparently excellent RV curves. Only 4 out of the 17 RN RV curves produced a plausible WD mass.  That is 76\% of the published RV WD masses for RNe are certainly bad, with systematic errors from 0.3--1.1 $M_{\odot}$.''  The evaluation of RV masses for novae was that these are not even a `bronze standard', but more like a `tin-foil standard'.}.  But the real condemnation is that a mass of 0.7$\pm$0.2 $M_{\odot}$ is impossible for a RN with a recurrence time as short as 12 years, so we know that there must be some large error in this RV claim.

Chomiuk et al. (2014) attribute the cool temperature ($\sim$45 eV) of the super-soft X-ray source to imply that $M_{\rm WD}$ is significantly below the Chandrasekhar mass, at $\sim$1 M$_{\odot}$.  But this attribution is only made as a non-quantitative comparison between T Pyx, RS Oph, and U Sco, with no physics, so this low mass estimate carries little confidence.  It is greatly more likely that the three-RN-comparison is complicated and inapplicable to the utterly unique T Pyx system rather than that the recurrence time scale can be as fast as 12 years with a 1 M$_{\odot}$ white dwarf.

Shara et al. (2018) model recurrent nova light curves, with $M_{\rm WD}$ being a function of the observed nova eruption duration and the average recurrence timescale.  This model has been validated for systems with high WD masses (Schaefer 2025e).  For T Pyx, they find a WD mass of 1.23 M$_{\odot}$.  Schaefer (2025e) measures the total error bar on this to be $\pm$0.15 M$_{\odot}$.

Schaefer (2025e) quantifies a number of nova properties that are correlated with $M_{\rm WD}$.  T Pyx has a P-class light curve, which requires a mass of $>$0.95 M$_{\odot}$.  T Pyx is of the He/N spectral class, which requires a mass of $>$1.15 M$_{\odot}$.  The correlation with the decline rate $t_3$ is superseded by the explicit calculation for T Pyx as made by Shara et al. (2018).  The FWHM of the H$\alpha$ emission line profile near maximum is 5350 km/s, which forces a high WD mass of $>$1.29 M$_{\odot}$ (see Figure 5 of Schaefer (2025e).

The observed time between eruptions places a very strong constraint on the possible mass of the white dwarf.  For RN in general with recurrence times of under 30 years, $M_{\rm WD}$$>$1.20 M$_{\odot}$.  This limit is forced by the physics of the triggering of nova explosion, and there is no real possibility of changing this limit.  This limit has been calculated by many workers (e.g., Truran and Livio 1986), and is nicely represented in the `Nomoto diagram' (see for example Figure 7 of Shen \& Bildsten 2009).  T Pyx itself has recurrence times of 12, 18, 24, 23, and 44 years.  For a 12 year recurrence time to be possible, $M_{\rm WD}$ must be near to the Chandrasekhar limit.  From the Nomoto diagram, in all cases, $M_{\rm WD}$ must be $>$1.31 $M_{\odot}$.  There is no way violate this strict limit.  So with strong confidence, we know that T Pyx has its WD mass from 1.31 to 1.40 M$_{\odot}$.

These physics calculations of the recurrence timescale can be inverted to give a measure of $M_{\rm WD}$ when the accretion rate is known.  For this, a convenient and authoritative Nomoto diagram is the plot of the accretion rate versus the WD mass as in Shen \& Bildsten (2009).  I here apply this for two eruptions:  First, the 1902 eruption had a recurrence time of 12 years, while the 1890--1902 observed accretion rate was close to the maximal value (i.e., just below the case of stable hydrogen burning) at near 2$\times$10$^{-7}$ M$_{\odot}$/year.  From the Nomoto plot, the WD mass must be near 1.33 M$_{\odot}$.  Second, for the 2011 eruption, the recurrence time was 44 years, while the accretion rate must be 3.8$\times$ lower than the maximal value (because the quiescent $B$ magnitude fell from its maximum of 13.95 mag before 1890 to an average $B$=15.40 from 1967--2011), to be near 0.5$\times$10$^{-7}$ M$_{\odot}$/year.  From the Nomoto plot, T Pyx must have a WD mass near 1.34 M$_{\odot}$.

So we have useful and reliable measures of 1.23$\pm$0.15, $>$1.29, $>$1.31, 1.33, and 1.34 M$_{\odot}$.  From this, I conclude that T Pyx has a WD mass of 1.33$\pm$0.02 M$_{\odot}$. 

\subsection{Companion Mass}

The only useful measure of $M_{\rm comp}$ comes from the Roche lobe size (0.202 $R_{\odot}$) plus the mass-radius relation.  For the reasonable mass-radius relation of main-sequence stars, the companion mass is 2.64 $M_{\odot}$ times $P$ in units of days (Equation 4.11 of Frank et al. 2002), or $M_{\rm comp}$=0.20 $M_{\odot}$.  The best mass comes from using the latest mass-radius relation for CV donor stars (see Figure 4 of Knigge et al. 2011), so we get $M_{\rm comp}$=0.16$\pm$0.02 $M_{\odot}$.

\subsection{Light Curve in Quiescence}

The $B$-band light curve in quiescence from {\it before} the 1890 eruption up until today shows stunning changes in the brightness.  In particular, T Pyx was at $B$=13.95 before the 1890 eruption, and it has been fading substantially, up until $B$=16.06 in 2025.  That is a fairly-steady secular decrease in light by a factor of 7.0$\times$.  So the scenario is that the 1866 ordinary nova eruption somehow ignited long lasting nuclear burning that irradiated the companion and forced an extreme accretion rate that then forced fast recurrent nova events.  (This extreme $\dot{M}$ in quiescence is orders of magnitude larger than for T Pyx before 1866, or for any other $P$=0.076 days CV.)  The secular fading from 1890--2025 shows that the positive feedback loop (the eruption making high-$\dot{M}$ making another eruption...) has a gain substantially less than unity.  This secular decrease in brightness (and in the accretion rate) is also seen in the time intervals between eruptions getting larger and larger, from 12 to 18 to 24 to 23 to 44 years.  We are watching the fading-out and slowing-down of the RN-phase of the T Pyx eruption cycle.  The rest of the eruption cycle is after the high-$\dot{M}$ phase is finished, with T Pyx back to its normal slow-accretion phase, building up gas on the WD for many millennia until the next classical nova kicks off another RN-phase.

The RN-phase of the T Pyx eruption cycle is charted by the $B$ band light curve.  Schaefer (2005) was the first to recognize this huge secular fading, with a plot up to 2005.  Schaefer et al. (2013) have a full plot up to the 2011 eruption in their Figure 1 and all the individual magnitudes are listed in their Table 2.  Schaefer (2023b) added on data up to 2022.  Now, in the current paper, I am making a variety of additions and improvements.

The entire light curve from 1890 to 1954 is from measures of the archival sky photographs stored at the Harvard College Observatory.  Importantly, these magnitudes are exactly in the modern $B$ magnitude system, because the spectral sensitivity of the plates is identical to that of the modern $B$ system\footnote{This is no accident, because the historical development of the modern system was explicitly designed to match the scale and zero of the Harvard North Polar Sequence first set up by Harvard Director W. Pickering.} and the comparison stars are exactly in the modern $B$ system\footnote{My original measures of the comparison star brightnesses (see Schaefer 2010) were explicitly calibrated against Landolt comparison stars.  Subsequently, comparison star magnitudes for all the AAVSO, APASS, and DASCH photometry are based on explicit calibration against Landolt comparison stars.}.  It is this identical magnitude system that ensures exacting continuity between the Harvard magnitudes and all modern $B$ photometry.  For the case of T Pyx (isolated, roughly 14th mag, good comparison stars nearby, multiple measures of each plate) the real photometric uncertainty is $\pm$0.10 mag.  While this is substantially poorer than that attainable by CCD photometry, this has negligible effect for the T Pyx light curve.  Part of the reason is that T Pyx flickers randomly with an amplitude of up to a quarter magnitude, so the total variance for any one magnitude is the addition in quadrature of the intrinsic flickering noise and the photometric uncertainty, so it matters little whether the photometric uncertainty is $\pm$0.10 mag, or $\pm$0.01 mag, or $\pm$0.001 mag, or $\pm$0.0001 mag.  With the flickering dominating, the important issue becomes how many measures of independent time intervals can be averaged together so as to beat down the flickering noise.  So, averaging together four Harvard plates will produce nearly the same total uncertainty in the T Pyx average brightness as from averaging together four nights of definitive photometry from A. Landolt himself.  So by averaging together multiple plates or multiple nights, the resultant light curve has the real magnitude uncertainty substantially smaller than the intrinsic flickering of T Pyx.  The resultant light curves are measures of T Pyx as averaged over the flickering.  The variability in the light curve is entirely dominated by the variability of T Pyx itself.  That is, the variations that we see are real measures of the changing T Pyx.

My original measures of the Harvard plates (Schaefer 2005; Schaefer et al. 2013) was made with selection of sky photographs (plates) that were easy to find and access.  Since then, the {\it Digital Access to a Sky Century @ Harvard} program\footnote{\url{https://dasch.cfa.harvard.edu/}} (DASCH, P. I. J. Grindlay) has completely digitized the Harvard plates and ran an impressive photometry package that measures the $B$ magnitudes for the stars on the plates.  Further, in 2020 and 2023, I have twice exhaustively sought out and measured {\it all} Harvard plates with T Pyx, and this includes many plates that had been missed by DASCH.  Now, the complete set of Harvard plates has typically 2--4 independent measures each, and this beats down the measurement errors\footnote{The measurement errors of the images on the plates are separate from the photometric errors for the images, arising for example from the randomness of grain development.  For ordinary good plates, the DASCH measurement errors are just as good as my very-experienced by-eye measures, as measured many times for many stars.  For the case of T Pyx, the typical measurement error is $\pm$0.10 mag, so multiple measures of the same plate can beat down the measurement-plus-photometric errors.}.  The Harvard plates were discontinued in 1954, and started up only in the late 1960s with the Damon patrol series, result in the notorious Menzel Gap\footnote{When D. Menzel became Director of the Harvard College Observatory, he killed the plate stack with active malevolence and devastation, by stopping the acquisition of new plates, the destruction of one-third of the existing plates, and the forced departure of all plate stack workers, all so that he could fund his own research (Hoffleit 2002).}.  After the 1967 eruption, T Pyx had become too faint to be visible on all-but-three of the Damon patrol plates.  The Menzel Gap can be seen in the final T Pyx light curve, with the measures only picking up when the A. Landolt started regular observing in 1969.  The result is 286 $B$-magnitudes from 1890 to 1954, and three more magnitudes from 1978--1983.

A substantial problem for getting one consistent long-term light curve is that few $B$-band measures have been made after the 2011 eruption.  Professional photometry of individual stars of interest has largely stopped, ceding the task to the world-class CCD photometry of the unpaid observers, as archived by the AAVSO's International Database.  And the vaunted sky surveys (ZTF, {\it TESS}, ASAS, ASAS-SN, {\it Gaia}, and the Rubin observatory) are not observing in the $B$-band, hence making problems with continuity with the historical record 1890--2011 that is in the $B$-band.  So we are left with the task of trying to extend the 1890--2011 record to the current day.  The only solution is to take some band relatively close to $B$, and convert it to $B$ with some constant color term.  The obvious band for conversion is the $V$ band\footnote{Use of the $g$ band is not a good idea because the coverage for 2011--2025 is spotty, and because many conflicting systems are in use, creating a `babel' of magnitude systems.}.  Fortunately, the $B-V$ color is constant throughout quiescence, and it is well-measured to be close to 0.14 mag.  So the AAVSO $V$-band light curve can be reliably converted to $B$-band by the simple offset of 0.14 mag.  So I am using the AAVSO $V$-band light curve, offset to $B$-band, for 25,296 magnitudes covering the time interval 1992--2025 (with exclusion of the interval around the 2011 eruption).

\begin{figure}
	\includegraphics[width=1.01\columnwidth]{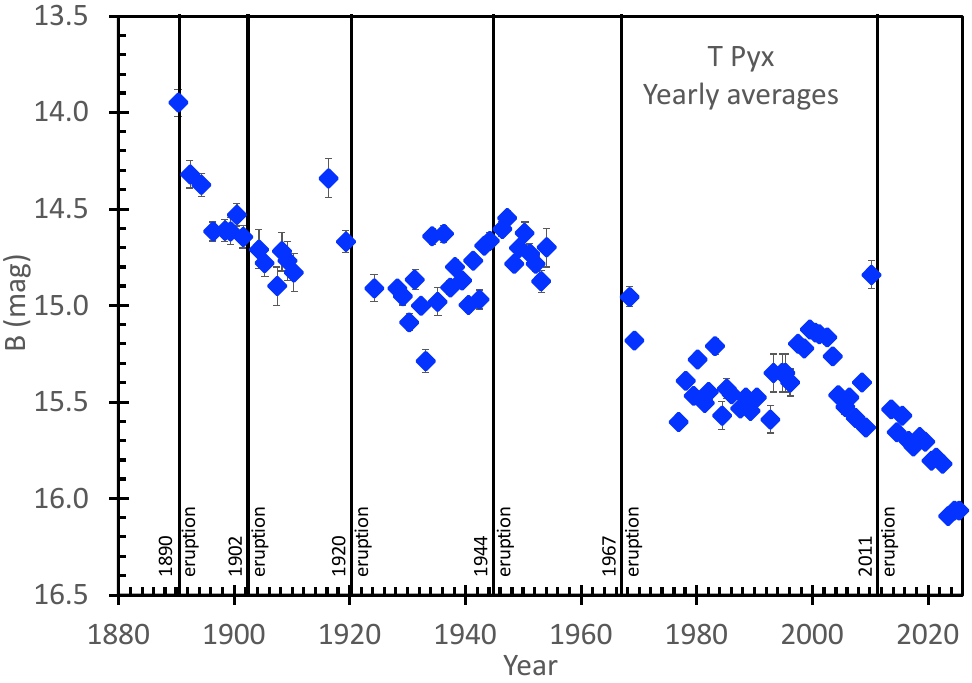}
    \caption{T Pyx yearly-averaged $B$ light curve.  This was constructed from 30,445 magnitudes, covering from {\it before} the 1890 eruption until near the end of 2025.  Importantly, we see T Pyx declines from $B$=13.95 in early 1890 to $B$=16.06 in 2025 (a factor of 7.0$\times$ change in flux).  This shows the slow decline of the extreme-high-accretion RN-phase (kicked off by the 1866 classical nova eruption) of the millennia-long T Pyx eruption cycle.  }
\end{figure}

\startlongtable
\begin{deluxetable}{lll}
\tablecaption{Yearly-averaged $B$ light curve of T Pyx 1890--2025}
\tablewidth{500pt}
\tabletypesize{\scriptsize}
\tablehead{
\colhead{$\langle Year \rangle$}   &   \colhead{$\langle B \rangle$}  &  \colhead{$\langle F_B \rangle$}    }         

\startdata
1890.34	&	13.95	$\pm$	0.07	&	6.61	\\
1892.35	&	14.32	$\pm$	0.07	&	4.77	\\
1894.35	&	14.38	$\pm$	0.06	&	4.48	\\
1896.31	&	14.62	$\pm$	0.05	&	3.62	\\
1898.37	&	14.61	$\pm$	0.06	&	3.65	\\
1899.33	&	14.62	$\pm$	0.07	&	3.60	\\
1900.45	&	14.53	$\pm$	0.06	&	3.90	\\
1901.51	&	14.64	$\pm$	0.06	&	3.50	\\
1904.27	&	14.71	$\pm$	0.10	&	3.28	\\
1905.31	&	14.78	$\pm$	0.07	&	3.08	\\
1907.37	&	14.90	$\pm$	0.10	&	2.75	\\
1908.24	&	14.72	$\pm$	0.10	&	3.25	\\
1909.24	&	14.77	$\pm$	0.10	&	3.10	\\
1910.35	&	14.83	$\pm$	0.10	&	2.94	\\
1916.27	&	14.34	$\pm$	0.10	&	4.61	\\
1919.30	&	14.67	$\pm$	0.06	&	3.41	\\
1924.23	&	14.91	$\pm$	0.07	&	2.74	\\
1928.26	&	14.91	$\pm$	0.03	&	2.74	\\
1929.13	&	14.95	$\pm$	0.04	&	2.63	\\
1930.32	&	15.09	$\pm$	0.04	&	2.33	\\
1931.28	&	14.87	$\pm$	0.05	&	2.87	\\
1932.25	&	15.00	$\pm$	0.03	&	2.54	\\
1933.10	&	15.29	$\pm$	0.06	&	1.95	\\
1934.26	&	14.64	$\pm$	0.04	&	3.53	\\
1935.21	&	14.98	$\pm$	0.07	&	2.56	\\
1936.22	&	14.63	$\pm$	0.04	&	3.58	\\
1937.31	&	14.91	$\pm$	0.04	&	2.79	\\
1938.22	&	14.80	$\pm$	0.04	&	3.05	\\
1939.41	&	14.87	$\pm$	0.03	&	2.86	\\
1940.51	&	15.00	$\pm$	0.04	&	2.53	\\
1941.27	&	14.77	$\pm$	0.03	&	3.19	\\
1942.37	&	14.97	$\pm$	0.05	&	2.60	\\
1943.30	&	14.69	$\pm$	0.04	&	3.39	\\
1944.19	&	14.66	$\pm$	0.04	&	3.48	\\
1946.35	&	14.60	$\pm$	0.04	&	3.65	\\
1947.24	&	14.55	$\pm$	0.04	&	3.87	\\
1948.41	&	14.78	$\pm$	0.03	&	3.11	\\
1949.24	&	14.70	$\pm$	0.03	&	3.35	\\
1950.26	&	14.63	$\pm$	0.06	&	3.68	\\
1951.14	&	14.73	$\pm$	0.05	&	3.24	\\
1952.19	&	14.78	$\pm$	0.04	&	3.09	\\
1953.10	&	14.87	$\pm$	0.06	&	2.87	\\
1954.02	&	14.70	$\pm$	0.10	&	3.31	\\
1968.38	&	14.96	$\pm$	0.05	&	2.65	\\
1969.23	&	15.18	$\pm$	0.03	&	2.15	\\
1976.90	&	15.61	$\pm$	0.03	&	1.44	\\
1978.15	&	15.39	$\pm$	0.03	&	1.81	\\
1979.48	&	15.47	$\pm$	0.03	&	1.64	\\
1980.11	&	15.28	$\pm$	0.02	&	1.94	\\
1981.40	&	15.51	$\pm$	0.02	&	1.58	\\
1982.08	&	15.45	$\pm$	0.04	&	1.66	\\
1983.16	&	15.21	$\pm$	0.04	&	2.19	\\
1984.36	&	15.57	$\pm$	0.07	&	1.49	\\
1985.17	&	15.43	$\pm$	0.05	&	1.69	\\
1986.08	&	15.46	$\pm$	0.02	&	1.65	\\
1987.52	&	15.54	$\pm$	0.02	&	1.54	\\
1988.43	&	15.48	$\pm$	0.02	&	1.62	\\
1989.25	&	15.55	$\pm$	0.02	&	1.52	\\
1990.34	&	15.48	$\pm$	0.02	&	1.62	\\
1992.78	&	15.59	$\pm$	0.07	&	1.46	\\
1993.26	&	15.35	$\pm$	0.10	&	1.82	\\
1994.80	&	15.35	$\pm$	0.10	&	1.82	\\
1995.28	&	15.35	$\pm$	0.10	&	1.82	\\
1996.16	&	15.40	$\pm$	0.07	&	1.74	\\
1997.50	&	15.20	$\pm$	0.03	&	2.09	\\
1998.59	&	15.23	$\pm$	0.03	&	2.06	\\
1999.55	&	15.13	$\pm$	0.02	&	2.24	\\
2000.55	&	15.14	$\pm$	0.02	&	2.21	\\
2001.22	&	15.15	$\pm$	0.04	&	2.19	\\
2002.56	&	15.17	$\pm$	0.04	&	2.16	\\
2003.56	&	15.27	$\pm$	0.02	&	1.97	\\
2004.46	&	15.46	$\pm$	0.02	&	1.65	\\
2005.69	&	15.53	$\pm$	0.04	&	1.56	\\
2006.44	&	15.48	$\pm$	0.02	&	1.63	\\
2007.45	&	15.58	$\pm$	0.02	&	1.48	\\
2008.52	&	15.40	$\pm$	0.02	&	1.74	\\
2009.31	&	15.63	$\pm$	0.02	&	1.42	\\
2010.20	&	14.84	$\pm$	0.07	&	2.91	\\
2013.58	&	15.54	$\pm$	0.02	&	1.55	\\
2014.56	&	15.66	$\pm$	0.02	&	1.40	\\
2015.50	&	15.57	$\pm$	0.02	&	1.54	\\
2016.69	&	15.70	$\pm$	0.02	&	1.32	\\
2017.51	&	15.73	$\pm$	0.02	&	1.29	\\
2018.57	&	15.68	$\pm$	0.02	&	1.34	\\
2019.55	&	15.71	$\pm$	0.02	&	1.31	\\
2020.60	&	15.81	$\pm$	0.02	&	1.20	\\
2021.44	&	15.79	$\pm$	0.02	&	1.21	\\
2022.53	&	15.82	$\pm$	0.02	&	1.18	\\
2023.44	&	16.09	$\pm$	0.02	&	0.92	\\
2024.58	&	16.07	$\pm$	0.02	&	0.94	\\
2025.43	&	16.06	$\pm$	0.02	&	0.96	\\
\enddata	
\end{deluxetable}

From all this, I have a T Pyx $B$ light curve with 30,445 magnitudes from 1890 to 2025.  I have averaged these together into yearly bins, with this beating down the usual flickering.  This binned light curve is listed in Table 1 and plotted in Figure 1.  This makes clear the steep secular decline from 1890 to 2025.  The magnitude faded from 13.95 to 16.06, with 2.11 magnitudes corresponding to a factor of 7.0$\times$ in flux.

The weighted average brightness throughout each of the inter-eruption intervals is tabulated in Table 2.  These magnitudes are also represented in units of the $B$-band flux where the unit is the flux from a $B$=16.00 star.  That is, $F_B$=10$^{-0.4*(B-16)}$.  The tabulated $\langle F_B \rangle$ is the average over the individual $F_B$ values included for each year.  The durations of the intervals are $\Delta T$ in years.

After a vast amount of work by hundreds of observers over the last 135 years, we have a good record of the brightness of T Pyx documenting the decline of the RN phase.  This light curve is the critical record of the changing accretion rate, fundamental for the analysis in this paper.

\begin{table*}
	\tablenum{2}
	\centering
	\caption{T Pyx light curve between eruptions}
	\begin{tabular}{llllll}
		\hline
		Years 	&	$\Delta T$  & $\langle B \rangle$ & $\langle F_B \rangle$ & $\langle \dot{M} \rangle$  &  $\langle \dot{M} \rangle$$\Delta T$    \\
		\hline
1883	 to 	1890.49	&	7.49	&	13.95	$\pm$	0.07	&	6.61	&	1.80$\times$10$^{-7}$	&	12.6$\times$10$^{-7}$	\\
1890.49	 to 	1902.35	&	11.86	&	14.54	$\pm$	0.02	&	3.93	&	1.07$\times$10$^{-7}$	&	14.1$\times$10$^{-7}$ ~*	\\
1902.35	 to 	1920.27	&	17.92	&	14.71	$\pm$	0.03	&	3.30	&	0.90$\times$10$^{-7}$	&	16.2$\times$10$^{-7}$	\\
1920.27	 to 	1944.90	&	24.63	&	14.87	$\pm$	0.01	&	2.85	&	0.78$\times$10$^{-7}$	&	19.2$\times$10$^{-7}$	\\
1944.90	 to 	1967.04	&	22.14	&	14.70	$\pm$	0.01	&	3.35	&	0.92$\times$10$^{-7}$	&	19.6$\times$10$^{-7}$ ~*	\\
1967.04	 to 	2011.36	&	44.32	&	15.40	$\pm$	0.01	&	1.83	&	0.50$\times$10$^{-7}$	&	22.2$\times$10$^{-7}$	\\
2011.36	 to 	2025.9	&	14.54	&	15.79	$\pm$	0.01	&	1.24	&	0.34$\times$10$^{-7}$	&	3.0$\times$10$^{-7}$ ~*	\\
		\hline
	\end{tabular}
	\\
	~~~~~~~~~~~~~~~~~* The values were calculated as a summation of a linear decline in magnitudes.
\end{table*}

\subsection{Accretion rate}

The accretion rate ($\dot{M}$) changes greatly from before 1866, to 1890, to 2025.  To know how much gas is falling onto the WD over its eruption cycle, we must map out the complex changes on a year by year basis.  I will map out these changes with four steps.  The first step is to collect and average the various independent estimates of the $\dot{M}$ for the years before the 2011 eruption.  The second step is to scale the accretion for all the history from 1890--2025 by the $B$ band flux, normalized to the 1967--2011 value.  This is making the good case that the brightness of T Pyx is proportional to the accretion rate, $\dot{M}$$\propto$$F_B$.  From the 1866 eruption to 1890, I can only presume that the accretion rate is fading to that of the pre-1890 eruption.  The third step is to estimate the accretion {\it before} the 1866 classical nova eruption as based on models and measures for the trigger mass for CVs of the known masses and periods.  The fourth step is to extrapolate from the 2011--2025 decline rate down to the quiescent level.  With these steps, the resultant $\dot{M}$ history from long before 1860 and up to the far future will be drawn up.  Figure 2 shows a schematic view of the accretion over the entire RN-phase of the long eruption cycle.  Values of accretion rates and the accreted masses for various intervals are collected into Table 3, for later comparison with the ejected masses.  This accretion history cannot be exact, but the accuracy will be adequate to answer the big questions. 

Schaefer (2005) and Schaefer et al. (2010) made the presumption that $\dot{M}$$\propto$$F_B^2$.  The basis for this was a standard calculation for an alpha-disk with T Pyx parameters, where the small disk size makes for a nearly blackbody SED, such that the changing accretion rate also changed the fraction of the disk flux in the $B$-band.   With this, the prediction was that T Pyx would erupt again only in the far future.  This presumption was refuted  when T Pyx had its long-awaited eruption in the year 2011.  This presumption was also refuted by Schaefer et al. (2013), where the SED was proven to certainly not be from any blackbody or disk, rather it was from an entirely non-thermal power law.  This SED does not have a shifting peak, so the flux will simply scale with the accretion rate.  With $\dot{M}$$\propto$$F_B$, the expected next eruption after 1967 was close to the year 2011.  Indeed, Schaefer et al. (2013) used the inter-eruption intervals and $\langle B \rangle$ data for T Pyx (like in Table 2) to measure the exponent, with the derived result being close to $\dot{M}$$\propto$$F_B$.  So for the estimates of $\dot{M}$ below, I will take the accretion rate to be simply proportional to the $F_B$.  To evaluate the constant of proportionality, I will collect independent measures of $\dot{M}$ and $\langle F_B \rangle$ for the time interval 1967--2011.

\begin{figure*}
	\includegraphics[width=2.1\columnwidth]{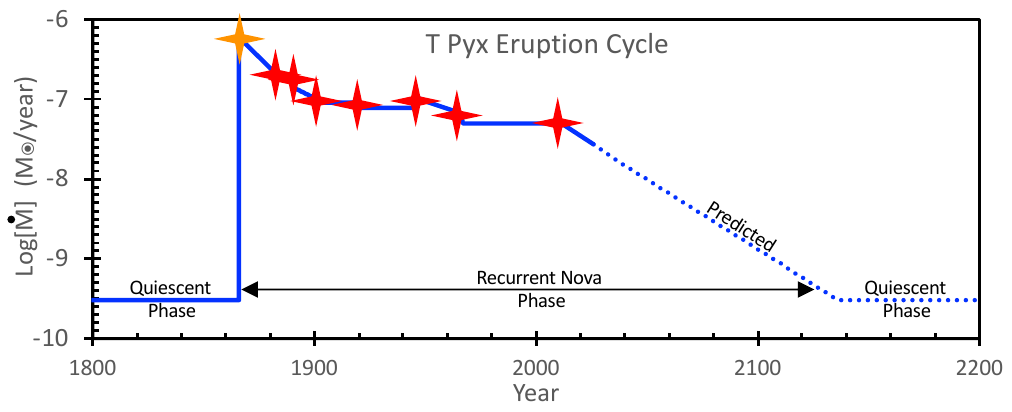}
    \caption{T Pyx accretion rate from 1800 to 2200.  The whole eruption cycle consists of an RN-phase alternating with a long quiescent phase.  The quiescent phase lasts perhaps 13,000 years, during which the WD accumulates gas slowly with a low-$\dot{M}$, as appropriate for T Pyx being under the Period Gap.  Around the year 1866, the gas accumulating on the WD reached the trigger mass, and the classical nova eruption ejects $\sim$10$^{-4.5}$ $M_{\odot}$ with velocities 500--715 km/s, as shown with the orange star.  For some reason (likely due to continued nuclear burning on the WD surface irradiating and bloating the companion star's atmosphere), the classical nova event drives extremely high $\dot{M}$.  The resultant high accretion rate starts a recurrent nova phase that lasts more than two centuries.  Throughout the RN-phase, we have been watching the accretion falling off (see Figure 1), as the feedback loop weakens.  This unique secular decline in $\dot{M}$ make the star's brightness fade and the time between RN eruptions (the red stars) increase from 12 years to 44 years.  The RN-phase has 6 observed RN events 1890--2011, plus another likely RN around 1883, indicated by the red stars.  For the future, presumably the $\dot{M}$ will keep fading at the 2011--2025 observed rate of 0.044 mag/year, with the extrapolated decline indicated by the dotted line.  After 2011, with any plausible decline rate, T Pyx will not accumulate the trigger mass until many millennia from now, so there will be no further RN eruptions in the current RN-phase.  While the uncertainty is large, T Pyx will return to it normal quiescent level sometime around the year 2137.  Then, it will take another 12,800 years or so to accumulate enough gas on the WD to trigger another classical nova eruption, to start the entire eruption cycle again.  }
\end{figure*}

\subsubsection{$\dot{M}$ from 1967--2011}

The accretion rate throughout the inter-eruption interval 1967--2011 is roughly constant, as demonstrated by the relatively flat $B$-band light curve (see Figure 1).  A wide variety of methods have been used to evaluate $\dot{M}$ for this time interval.  Here, I present four methods.    (Schaefer et al. 2013 discusses two more methods that ultimately failed.)

{\it Method 1.}  The light from T Pyx is essentially all from the accretion.  The accretion luminosity from the accretion disk is $0.5GM_{\rm WD} \dot{M}/R_{\rm comp}$.  This equation can be used to calculate $\dot{M}$ from some measure of the accretion luminosity.  Selvelli et al. (2003) measured the ultraviolet luminosity with their {\it IUE} spectra and with a variety of corrections, they ended up with $\dot{M}$=(0.22--0.46)$\times$10$^{-7}$ M$_{\odot}$/year.  Selvelli et al. (2008) refined this analysis and reported $\dot{M}$$\sim$0.11$\times$10$^{-7}$ M$_{\odot}$/year.  Selvelli et al. (2010) present the same estimate with quoted error bars as $\dot{M}$$\sim$(0.11$\pm$0.03)$\times$10$^{-7}$ M$_{\odot}$/year.   Patterson et al. (2017) used the same relation to calculate $\dot{M}$=1.1$\times$10$^{-7}$ M$_{\odot}$/year, which is a factor of 10$\times$ larger than from the Selvelli group.  So the uncertainty from this traditional method appears to be a factor-of-ten.  But the situation is worse than that, because these calculations assume that all the light is coming from an accretion disk, whereas such is impossible because the SED is a nearly-perfect power law from the middle-infrared to the middle-ultraviolet.  The SED certainly does not display the required shape of a disk, so there must be a substantial added luminosity from some hot non-thermal source.  Presumably, this extra light is somehow associated with the hot source around the WD that is driving the high accretion.  The disk luminosity is then less-then, and maybe greatly less-than, the values adopted by the Selvelli and the Patterson papers.  This would make the actual accretion rate smaller than calculated.  So these measures should be taken to say that the accretion rate is $<$(0.1--1.1)$\times$10$^{-7}$ M$_{\odot}$/year.

{\it Method 2a.}   The thermonuclear runaway is triggered when the accumulated gas reaches some trigger mass.  The average $\dot{M}$ from 1967--2011 will equal the trigger mass divided by 44 years.  The trigger mass for various cases can be presented in a `Nomoto plot' (for example, see Figure 7 of Shen \& Bildsten 2009), which is a plot of $\dot{M}$ versus $M_{\rm WD}$.   From the reliable physics of the triggering of the runaway explosion, both $M_{\rm trigger}$ and the recurrence timescale can be plotted as curves on the Nomoto plot.  The case for T Pyx in 2011 is for a recurrence interval of 44 years and the WD mass of 1.33 $M_{\odot}$, with these two constraints crossing at one point in the Nomoto diagram.  For this point, we can read off the average accretion and the trigger mass.  With this, the indicated accretion rate is close to 0.5$\times$10$^{-7}$ M$_{\odot}$/year, and the trigger mass is near 22$\times$10$^{-7}$ M$_{\odot}$.  So by this version of Method 2, $\dot{M}$ before the 2011 eruption is 0.5$\times$10$^{-7}$ M$_{\odot}$/year.

{\it Method 2b.}   The trigger mass can be independently gotten from the model calculations of Yaron et al. (2005).  In their Tables 2 and 3, for $M_{\rm WD}$=1.25 $M_{\odot}$, for linear interpolation in $\log[M_{\odot}]$, they calculate the trigger mass to be 24$\times$10$^{-7}$ M$_{\odot}$ and an accretion rate of 0.54$\times$10$^{-7}$ M$_{\odot}$/year.  This result is close to that for Method 2a.  (That is, the physics is straight forward and reliable.)  From this, I conclude $\dot{M}$=0.5$\times$10$^{-7}$ M$_{\odot}$/year, for the 1967--2011 interval.

{\it Method 3.}  Before the 1890 eruption, I have two Harvard plates with the average quiescent magnitude of $B$=13.95.  The $\dot{M}$ before the 1890 eruption must be below the threshold for steady hydrogen burning and likely fairly close to the threshold.  This limit can be seen in the Nomoto plot.  For a 1.33 $M_{\odot}$ WD, the threshold is 2$\times$10$^{-7}$ M$_{\odot}$.  In the 1967--2011 interval, the average $B$ magnitude is 15.40.  The ratio of the flux before 1890 and the 1967--2011 flux is 3.6$\times$.  This ratio is also confirmed by the ratio of the longest recurrence time (44 years) to the shortest recurrence time (12 years), for a factor of 3.7$\times$.  The observed flux is going to be proportional to $\dot{M}$, whether the flux is from an accretion disk or from nuclear burning on the WD.  So the accretion rate in 1967--2011 is 3.6$\times$ lower than $\lesssim$2.0$\times$10$^{-7}$ M$_{\odot}$/year.  So the accretion rate before the 2011 eruption was $\lesssim$0.56$\times$10$^{-7}$ M$_{\odot}$/year.  This method is the one that I consider to be the most reliable of all.  The reason is that the only assumptions are that the $B$-band brightness is proportional to $\dot{M}$ and that the steady-hydrogen-burning threshold has not been violated leading up to either the 1890 or 1902 nova eruptions and is near 2.0$\times$10$^{-7}$ M$_{\odot}$/year.  For both assumptions, I see no realistic possibility of any significant deviation, so this derived accretion rate is reliable, even if only a limit.

{\it Method 4.}   From 1985 to 2011, I measure the $\dot{P}$ to be ($+$6.49$\pm$0.07)$\times$10$^{-10}$, in dimensionless units.  The ordinary mass transfer in the binary produces a steady period change,
\begin{equation}
\dot{P}_{\rm mt} = 3 P (1+q) \frac{\dot{M}}{M_{\rm comp}},
\end{equation}
where $q$ is the mass ratio ($M_{\rm comp}/M_{\rm WD}$).  This can be turned around to derive the $\dot{M}$.  With the masses and period of T Pyx, I calculate $\dot{M}$=1.5$\times$10$^{-7}$ M$_{\odot}$/year.  This is immediately below the threshold for steady hydrogen burning and is certainly too high of an estimate.  (That is, if this accretion rate is correct, then the rate a century earlier would be so high as to make any nova runaway impossible.)  The problem undoubtedly lies in that the simple mass transfer effect is relatively small compared to other effects (from unknown mechanisms) that dominate the observed $\dot{P}$.  That is, the measured $\dot{P}$ from the $O-C$ curve is not a good measure of the mass-transfer $\dot{P}_{\rm mt}$.  Indeed, Schaefer (2022a; 2024) measured $\dot{P}$ for many novae and CVs, always finding that the observed values are greatly different from that predicted by mass transfer alone, with these varying by orders of magnitude, both positive and negative.  Unknown mechanisms are dominating the $\dot{P}$ for real CVs, and this is actually a major problem for all of binary evolution (Schaefer 2024; 2025b).  So it is not surprising for the $\dot{M}$ from $\dot{P}_{\rm mt}$ having errors of several orders-of-magnitude.  That is, this fourth method to derive the accretion rate is a failure because the observed $\dot{P}$ is not a useful measure of $\dot{P}_{\rm mt}$.

Only the first three methods here provide measures of any usable reliability.  The three measures are $<$(0.1--1.1), 0.5, and $\lesssim$0.56, all in units of 10$^{-7}$ M$_{\odot}$/year.  These are all consistent with $\dot{M}$=0.5$\times$10$^{-7}$ M$_{\odot}$/year.  The constraints from the second and third method are each fairly exacting and hard to get around, and they are only in agreement close to 0.5$\times$10$^{-7}$ M$_{\odot}$/year.  I evaluate the error bar to be 20\%.  So $\dot{M}$ averaged 1967--2011 is (0.5$\pm$0.1)$\times$10$^{-7}$ M$_{\odot}$/year.  With this average accretion rate working for 44.32 years, the mass accreted in this inter-eruption interval is $\langle \dot{M} \rangle$$\Delta T$=22.2$\times$10$^{-7}$ M$_{\odot}$ (see Tables 2 and 3).

\subsubsection{$\dot{M}$ from 1883 to 2025}

For the light from the accretion disk, the brightness is linearly proportional to $\dot{M}$.  For the light from any nuclear burning around the WD, the brightness is linearly proportional to $\dot{M}$.  So from any combination of light sources, the observed brightness of T Pyx will scale up and down with changes in the accretion rate.  That is, the $B$-band light curve in quiescence from before the 1890 eruption up until today provides a nice measure that is proportional to the $\dot{M}$.  The constant of proportionality can be set by the average $\dot{M}$ from 1967--2011.  So a $B$ magnitude of  15.40$\pm$0.01 (with $F_B$=1.83$\pm$0.02) corresponds to $\dot{M}$=(0.5$\pm$0.1)$\times$10$^{-7}$ M$_{\odot}$/year.  With this, we can scale the light curve to get $\dot{M}$ from before the 1890 eruption up until today.

Between the 1944 and 1967 eruptions, we only have the light curve from 1944--1954, due to the Menzel Gap.  If the $B$-band brightness was constant, $\langle B \rangle$=14.70$\pm$0.01, which corresponds to $\langle \dot{M} \rangle$ of 0.92$\times$10$^{-7}$ M$_{\odot}$/year, and a total mass of accreted gas to be $\langle \dot{M} \rangle$$\Delta T$ equal to 20.3$\times$10$^{-7}$ M$_{\odot}$.  This is for the case of a flat light curve.  But, given the lack of information during the Menzel Gap, it is possible that T Pyx had some sort of a linear decline from 1944 to 1967.  A linear decline in {\it magnitudes} of 0.024 mag/year corresponds to a nonlinear decline in $F_B$ and in the corresponding $\dot{M}$.  For each year and partial year between 1944.90 to 1967.04, with a linear decline in magnitudes, the $F_B$ is converted to $\dot{M}$, and the mass accreted in each interval is calculated.  The sum of the accreted mass for the intervals is 18.9$\times$10$^{-7}$ M$_{\odot}$.  We do not have the information to know whether T Pyx followed a flat light curve or a declining light curve.  Nevertheless, the difference between a flat and a declining light curve is fairly small.  So I will take the average, letting the spread represent this uncertainty, so the accreted mass is (19.6$\pm$0.7)$\pm$0.07)$\times$10$^{-7}$ M$_{\odot}$.  This accreted mass is quoted into Tables 2 and 3, so that we can see all the various contributions to $M_{\rm accreted}$ over the full millennia-long eruption cycle of T Pyx.

Between the 1920 and 1944 eruptions, for 24.62 years, T Pyx was relatively constant, with $\langle B \rangle$=14.87$\pm$0.01 (see Table 2).  With scaling from the 1967--2011 interval, this corresponds to $\langle \dot{M} \rangle$ of 0.78$\times$10$^{-7}$ M$_{\odot}$/year.  This makes for a total mass of accreted gas to be $\langle \dot{M} \rangle$$\Delta T$ equal to 19.2$\times$10$^{-7}$ M$_{\odot}$.  This accreted mass is passed into Table 3, so as to tally up the various contributions to $M_{\rm accreted}$.

Between the 1902 and 1920 eruptions, for 17.92 years, $\langle B \rangle$=14.71$\pm$0.03.  This scales to $\langle \dot{M} \rangle$ of 0.90 $\times$10$^{-7}$ M$_{\odot}$/year and $\langle \dot{M} \rangle$$\Delta T$=16.2$\times$10$^{-7}$ M$_{\odot}$. 

Between the 1890 and 1902 eruptions, for 11.86 years, the average $B$ is 14.54$\pm$0.02, for $\langle \dot{M} \rangle$ of 1.07 $\times$10$^{-7}$ M$_{\odot}$/year and $\langle \dot{M} \rangle$$\Delta T$=12.7$\times$10$^{-7}$ M$_{\odot}$.  The 1890--1902 light curve looks to have a linear decline in {\it magnitudes}.  The decline in flux and in $\dot{M}$ was not linear, so the average $\dot{M}$ will be slightly different from the calibration based only on $\langle B \rangle$.  So I have calculated $F_B$ from the linear $B$ for each year and fraction of year, calculated $\dot{M}$, then summed up $\dot{M}$ times the interval duration, so as to get the total accreted mass.  With this, the mass accreted between eruptions was 14.1$\times$10$^{-7}$ M$_{\odot}$.  This value is to be preferred as a more accurate representation of the observed light curve.

Before the 1890 eruption, the primary evidence is the two Harvard plates with $B$=13.95$\pm$0.07.  This is 0.59 mag brighter than during the 1890--1902 interval.  This implies an accretion rate of 1.80 $\times$10$^{-7}$ M$_{\odot}$/year.  This accretion rate is slightly below the threshold for steady hydrogen burning, as it must be for there to be a nova eruption in 1890.  The duration of the inter-eruption interval before 1890 must be close to 11.86 years times the ratio of $F_B$, which is to say that the 1890 per-eruption interval was near 7.0 years.  This put the prior eruption close to the middle of 1883.  With this, $\langle \dot{M} \rangle$$\Delta T$=12.6$\times$10$^{-7}$ M$_{\odot}$.

After the 2011 eruption, T Pyx shows a roughly linear decline in magnitudes.  The average $B$ magnitude from 2011--2025 is 15.79$\pm$0.01, with a corresponding accretion rate of 0.34$\times$10$^{-7}$ M$_{\odot}$/year and a total accreted mass of 5.0$\times$10$^{-7}$ M$_{\odot}$.  A better calculation is to use the linear trend in magnitudes (0.044 mag/year), convert to $F_B$ and $\dot{M}$ for one year intervals.  The sum of the masses accreted during the intervals is 3.0$\times$10$^{-7}$ M$_{\odot}$ from 2011.36 to 2026.0.

\subsubsection{$\dot{M}$ from 1866 to 1883}

The $\sim$1866 classical nova eruption kick-started the extreme accretion.  By sometime around 1876 or 1883, the $\dot{M}$ must have fallen to below the critical threshold for steady hydrogen burning (2.0$\times$10$^{-7}$ M$_{\odot}$/year), or else there could be no 1890 eruption.  Perhaps the $\dot{M}$ fell below the steady hydrogen burning threshold around 1883, and the WD spent the next 7 or so years accumulating unburnt hydrogen to initiate the first RN eruption in 1890.  Or perhaps the $\dot{M}$ fell below the threshold some time around 1876, and then spent the time up until 1883 accumulating unburnt hydrogen to burn during the first RN eruption.  The apparent scenario is that the 1866 classical nova eruption kicked $\dot{M}$ up to some value like 6.0$\times$10$^{-7}$ M$_{\odot}$/year (the top of the stable hydrogen burning region), which then declined to 1.8$\times$10$^{-7}$ M$_{\odot}$/year (just below the bottom of the stable hydrogen burning region) around the year 1876 or 1883.  Such would be an average accretion rate of 4$\times$10$^{-7}$ M$_{\odot}$/year for roughly 17 years from 1866--1883.  With this, the accreted mass is 68.0$\times$10$^{-7}$ M$_{\odot}$.  Round about the year 1883 or earlier, the accretion fell below the steady burning regime, and unburnt hydrogen started accumulating, leading to the 1890 eruption.

\subsubsection{$\dot{M}$ from before 1866}

Around 1866, T Pyx suffered a normal classical nova eruption, with a shell of $\sim$10$^{-4.5}$ $M_{\odot}$ and expansion velocity of 500--715 km/s.  This was a completely different kind of event from the 6 RN events from 1890--2011.  The difference is that the RN events are fed by an extremely high accretion rate, as observed for the extremely high absolute magnitude for the system (see Figure 1).  The reason for the high $\dot{M}$ from 1866 to now must be caused by the WD somehow making the heavy irradiation of the close-in companion star driving the mass loss, but the detailed mechanism is not known.  Before 1866, T Pyx was just a normal CV with a short $P$ and a high $M_{\rm WD}$.  Before 1866, T Pyx could only have been accreting at a rate normal for CVs of $P$=0.076 days.  We cannot know this average $\dot{M}$ with any exactitude, nor can we know for how long T Pyx was accumulating gas over the previous long-quiet part of its cycle of eruptions, $\Delta T$.  But we can know the product $\langle \dot{M} \rangle$$\Delta T$, because that is the trigger mass for the classical nova event, and that we have modestly good estimates from nova physics.  Fortunately, the trigger mass is the quantity that we need for working out the mass budget of T Pyx over a complete eruption cycle.

Dubus et al. (2018) lists and plots $\dot{M}$ for 110 CVs of all types and for all periods.  This shows us the median and range for CVs with periods near 0.076 days.  We see that the median accretion rate is near 0.003$\times$10$^{-7}$ M$_{\odot}$/year, so this is the best estimate for T Pyx before 1866.  This best estimate $\dot{M}$ corresponds to $B$=21.0 mag.  The extreme range of accretion is from 0.000015 to 0.008 times10$^{-7}$ M$_{\odot}$/year.  

Yaron et al. (2005) calculates the trigger mass as a function of $\dot{M}$ for a range of appropriate cases.  They label it as the mass accreted before the eruption, $m_{\rm acc}$, which is exactly what is needed.  From their Table 2, for WD masses of 1.25 $M_{\odot}$, with linear interpolation in $\log[\dot{M}]$, we can work out their trigger masses, for the range of conditions of WD temperatures.  For the best estimate accretion rate of 0.003$\times$10$^{-7}$ M$_{\odot}$/year and the middle temperature, the trigger mass is 48$\times$10$^{-7}$ M$_{\odot}$/year.  This gives a recurrence time of 16,000 years.  For T Pyx with its RN-phase and substantial accretion left unburnt when the system reaches quiescence, the cycle time will be shorter than 16,000 years, which I calculate to be 13,000 years.  For the best estimate accretion rate and for the range of temperature, the trigger mass ranges (27--56)$\times$10$^{-7}$ M$_{\odot}$/year, for times since the last eruption of 9000 years to 19,000 years.  For the middle temperature, over the extreme range of accretion rates, the trigger mass ranges (20--51)$\times$10$^{-7}$ M$_{\odot}$/year, while the time since the last eruption ranges from 6400 years to 1,432,000 years.  So for T Pyx in 1866, the best estimate of the trigger mass is 48$\times$10$^{-7}$ M$_{\odot}$/year, with an extreme range of (20--56)$\times$10$^{-7}$ M$_{\odot}$/year.

\subsubsection{$\dot{M}$ after 2026}

Presumably, T Pyx will keep fading back to its baseline level after the end of the RN phase.  The exact baseline level is not known, but the best estimate is for the quiescent level at $\dot{M}$=0.003$\times$10$^{-7}$ M$_{\odot}$/year (see the previous Section).  This corresponds to $F_B$=0.010 and $B$=21.0.  Presumably, the fade rate will be an extension of the 2011--2025 fade rate of 0.044 mag/year.  To fade from $B$=16.06 in 2025 to $B$=21.0 at the rate of 0.044 mag/year, T Pyx will have to wait 112 years.  That is, at this rate, T Pyx will fade to quiescence around the year 2137.  The entire RN-phase will last from roughly 1866 to 2137, or 271 years.  Formally, for 2026--2137, the year-by-year summation of the accretion will be 6.5$\times$10$^{-7}$ M$_{\odot}$.  The uncertainties in this extrapolation are large and difficult to quantify.  Fortunately, the mass accreted is relatively small, so these uncertainties are of little consequence.

For the years 2011--2137, T Pyx will have accreted a total of 9.6$\times$10$^{-7}$ M$_{\odot}$.  This is to be compared to the trigger mass leading up to the 2011 eruption of 22.2$\times$10$^{-7}$ M$_{\odot}$ and the trigger mass leading up to the 1866 classical nova eruption of (20--56)$\times$10$^{-7}$ M$_{\odot}$.  This means that T Pyx has no real chance of erupting before the end of the RN-phase of the long eruption cycle.

For the best estimate trigger mass of 48$\times$10$^{-7}$ M$_{\odot}$, T Pyx will have to accumulate 38$\times$10$^{-7}$ M$_{\odot}$ more gas during quiescence.  At an accretion rate of 0.003$\times$10$^{-7}$ M$_{\odot}$/year, it will take 12,800 years until the next classical nova eruption starts the whole cycle over again.  Formally, this would make the next eruption in the year 12,800+2137=14,937 AD.  Clearly, these are huge uncertainties in this date.  Fortunately, for the purposes at hand, calculating $M_{\rm accreted}$ over the entire eruption cycle, the uncertainties are modest, because the mass is actually taken from the trigger mass for T Pyx in the quiescent state.

\subsection{Orbital Period}

The orbital period is the one most important property of any CV.  Szkody \& Feinswog (1988) saw T Pyx fading-then-brightening-then-fading over a 130 minute $J$-band photometry run, and reported an approximate period of 0.069 days.  Barrera \& Vogt (1989) reported a radial velocity curve with a period of 0.1433 days and amplitude of 29.1$\pm$14.9 km/s.  Schaefer (1990) used 372 $B$ magnitudes to report a photometric periodicity of 0.0991 days, although this was noted to be just the highest peak in the Fourier transform with prominent daily aliases (including what is now know to be the true $P$).  Schaefer et al. (1992) collected 1713 magnitudes and resolved the alias problem to determine a period of 0.07616$\pm$0.00017 days.  Patterson et al. (1998) carried out extensive photometry in 1996--1997, confirming the significance and stability of the period at 0.076227 days, but were confused about the nature of this periodicity because they also detected a transient signal with periodicity of 0.1098 days, as well as a `complex signal' at 1.25 days in February 1996.  (These non-orbital signals have never been seen before or since, especially to very deep limits within the excellent {\it TESS} light curves.)  Uthas, Knigge, \& Steeghs (2010) confirmed the orbital period at 0.076229 days with a radial velocity curve, removing all prior doubts about period-doubling and superhumps.

Livio (1991) noted that T Pyx should be erupting soon, so observers had an opportunity to measure a very accurate {\it pre}-eruption period ($P_{\rm pre}$), followed by a {\it post}-eruption period  ($P_{\rm post}$) measured in later years.  Livio also sketched out basic theory for orbital changes across a nova eruption, and this is largely the modern theory in Section 3.3.  Livio realized that an experimental measure of $\Delta P$ ($P_{\rm post}$-$P_{\rm pre}$) would determine the change in size of the orbit.  This was building on the result of Schaefer \& Patterson (1983) in which the orbital period change of classical nova BT Mon was accurately measured (taking advantage of the Harvard archival plates to fortuitously measure the pre-eruption period), with the measure of ejected mass at 3$\times$10$^{-5}$ M$_{\odot}$.  This was part of my original motivation for starting the career-long program of measuring $\Delta P$ in recurrent nova and novae.  The realization was that an accurate $\Delta P$ would produce a measure of the nova's ejected mass ($M_{\rm ejecta}$) and provide an answer to whether the popular supernova-progenitor candidates could actually work.  The task of measuring $\Delta P$ was known then to require many decades with relentless observing.  The selection of recurrent novae as targets made possible the measurement of the period {\it before} the upcoming eruption, confident in the inevitability that the eruption was coming.  This paper is the culmination of my career-long program for T Pyx, answering the question as to whether it can possibly become a Type Ia supernova.

\subsection{Times of photometric minima}

T Pyx displays  perfectly periodic photometric modulations, with period $P$=0.076227 days.  The light curve is typical for CVs, with a broad minimum and a relatively flat maximum (see Figure 3).  At first glance, it looks like a primary eclipse, and we know from the radial velocity curve (Uthas, Knigge, \& Steeghs 2010) that the minimum is at the time when we could expect an eclipse of the accretion disk by the small companion star.  But the observed minimum has a total duration nearly 0.5 in phase, and so this minimum cannot be an eclipse.  Rather, the modulation must arise from the beaming pattern of the hot light from around the WD, from the disk and hot spot, and irradiation of the inner hemisphere of the companion.  Whatever the combination of hot light, the photometric periodicity is coherent over decades and so must be tied exactly to the binary orbit.  As such, timings of the photometric minima are a measure of the orbital phase.

I started measuring T Pyx light curves back in 1987, using telescopes at Cerro Tololo and Kitt Peak observatories (Schaefer et al. 1992), seeking the orbital period as a measure of $P_{\rm pre}$.  A. Landolt had taken time series on 60 nights from 1967 to 1980 at the same observatories.  And from the literature and private communications, further time series were collected from 1966--1990 (Schaefer et al. 1992).  Then, J. Patterson, J. Kemp, plus many of the observers with the Center for Backyard Astrophysics observed an impressive set of minima timings from 1995--2016 (Patterson et al. 2017).  After 2016, many observers have archived time series photometry into the International Database of the  American Association of Variable Star Observers (AAVSO), with my calculating 31 times of minimum from 2016 to 2022 (Schaefer 2023b).  And, I pulled out a single time of minimum for the $\sim$25 days of nearly gap-free light curves from each of {\it TESS} Sector 8 (February 2019) and Sector 35 (February 2021), both with excellent timing accuracy as an average of over 300 orbits.  In all, I have collected 129 times of minimum for T Pyx from 1986 to 2022 (Schaefer 2023b).

\begin{figure}
	\includegraphics[width=1.0
	\columnwidth]{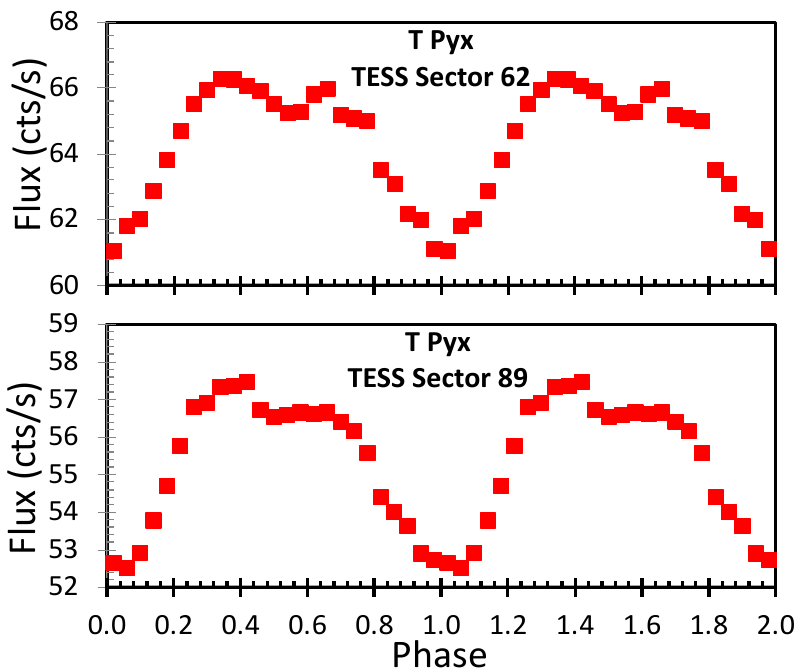}
    \caption{{\it TESS} folded and phase-averaged light curves.  The light curves show an ordinary CV, with the broad minimum at the time when the radial velocity curve has the companion star exactly in front of the WD.  Nevertheless, the primary minimum cannot be from an eclipse, because the total duration of the minimum is nearly 0.5 in phase.  }
\end{figure}

Further {\it TESS} Sectors have become available, for Sector 62 (February 2023) and Sector 89 (February 2025).  Sector 62 has 24 days of nearly gap-free photometry with 16105 measured fluxes of 120 second time resolution.  Sector 89 has 28 days of nearly gap-free photometry with 20181 fluxes of 120 second time resolution.  I have used the mission standard light curves labeled as `SPOC'.  The {\it TESS} times of mid-exposure are in Barycentric Julian Days (BJD, with Dynamical Time), which has only a negligibly small difference from the Heliocentric Julian Days (HJD, with UT), as used for all other minimum times.  The photometric error in the fluxes (expressed in counts per second) are $\pm$6 for both Sectors.  The measured flux includes substantial contribution from ordinary sky light and faint background stars within the {\it TESS} pixel size of 21$\times$21 arc-seconds.  T Pyx shows its usual substantial flickering, so the folded light curves show a scatter larger than the periodic modulation, but there is so much data that the phase-averaged light curves have averaged out all the flickering and all the Poisson measurement errors.  

The folded and phase-averaged light curves are shown in Figure 3.  These are comparable to the folded light curves shown in Figure 4 of Patterson et al. (2017).  For CVs, it is common enough to have small differences where the peak at phase 0.25 is somewhat brighter than the peak at 0.75.  For Sector 89 (but not for Sector 62), the primary minimum has a slower ingress than egress.  The particular light curve shape changes over time, certainly related to ordinary variations in the structure of the accretion and hot region around the WD.

The times of minimum are derived from a chi-square fit from the data versus a periodic template.  For timing purposes, the important point is that the template is symmetric around the minimum, with a nearly parabolic shape.  To avoid contribution to the minimum time from the variable and asymmetric maximum, the template is flat from phases 0.25--0.75.  With the period, amplitude, and average for the template fixed, the time of minimum is varied.  The particular cycle of the minimum is chosen from a time in the middle of the observations.  The best fit is determined by minimizing the chi-square, with the one-sigma error bars determined from the region of parameter space over which the  chi-square is within 1.00 of the minimum.  Importantly, the many other minimum times are defined by effectively equivalent calculations.

My two new new times of minima are BJD 2460002.05856$\pm$0.00031 and 2460732.01041$\pm$0.00026.  These are shown in the upper-right corner of Figure 4.  These times fit perfectly onto the best fit $O-C$ curve from Schaefer (2023b).  The best fit parameters from 2023 are not sensibly changed.

\begin{figure}
	\includegraphics[width=1.0
	\columnwidth]{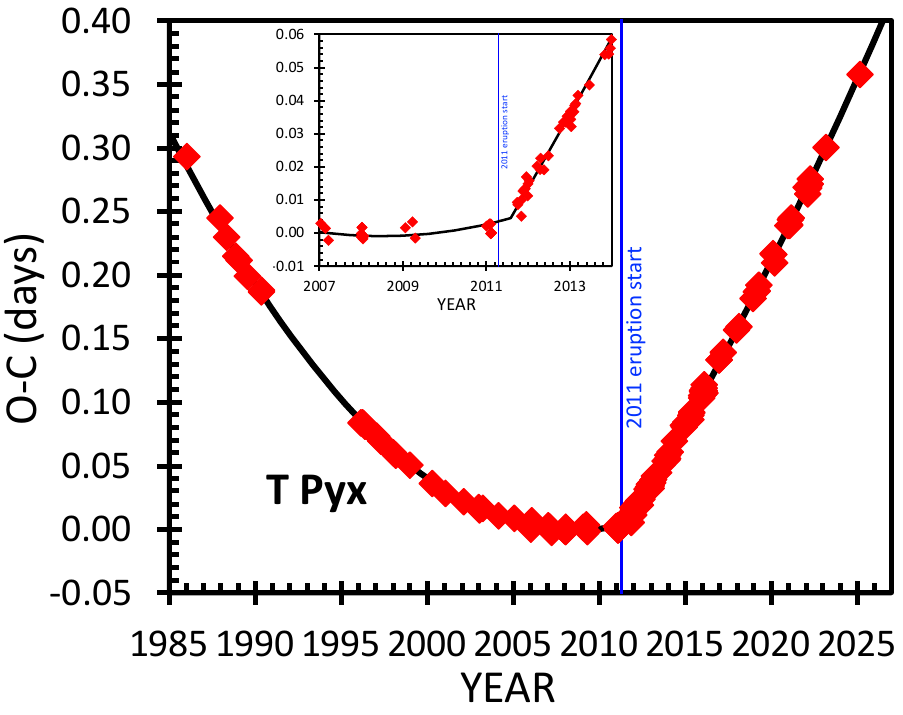}
    \caption{$O-C$ curve for T Pyx.  This plot has 131 times of minimum (red diamonds) from 1986 to 2025, with the ephemeris having period 0.07622916 days and epoch HJD 2455665.9962.  The best-fit broken parabola is shown with the black curve, which is a remarkably good match to the observed times.  The main point of this figure is that T Pyx had a sharp upward kink at the time of the 2011 eruption (denoted by the blue vertical line), which shows that the orbital period {\it increased} by $+$50.3$\pm$7.9 parts-per-million.  This is the long and hard result that provides a strict lower limit on the mass ejected during the 2011 eruption.  A further important point is that the kink is at 108.4$\pm$12.6 days after the start of the eruption (see inset).  This means that the median ejection was late in the eruption, and the mass ejection continued for hundreds of days.  This disproves the old standard idea that the ejections are some optically thick wind blasted off by the original thermonuclear runaway, and rather that the ejections are driven by the companion orbiting within the hot envelope surrounding the WD.  A third important point from this figure is that the very well-measured $\dot{P}$ (shown by the parabolic curvature) is large-and-positive before 2011, and much smaller and positive after the 2011 eruption.  {\it Something} about the nova eruption changed the state of the binary for at least the next 15 years of quiescence, with the obvious mechanism being that the large mass ejection enlarged the binary orbit so that the Roche lobe rises in the companion's atmosphere, which lowers the $\dot{M}$ and decreases the $\dot{P}$.  A fourth point is that the $\dot{P}$ values are {\it positive}, and this is impossible in the Magnetic Braking Model for CV evolution (Knigge et al. 2011), with this being just one of many counterexamples (Schaefer 2024; 2025b).  A fifth point is that the T Pyx evolution has the period increasing from beginning-to-end, contrary to the pervasive idea that all CVs evolve from long-$P$ to short-$P$.  This case is just one counterexample out of many, where roughly half of all CVs and XRBs have their periods {\it increasing} over the entire observed range (Schaefer 2024; 2025b).  This sampling of 77 systems disproves the ubiquitous ideal that all CVs and XRBs evolve from long-period to short-period.}
\end{figure}

\subsection{$\Delta P$ across 2011 eruption}

For this paper, the important point is that the $O-C$ curve provides an accurate measure of the $\Delta P$, and hence a strong limit on $M_{\rm ejecta}$.  To be specific, the period immediately before the 2011 eruption was $P_{\rm pre}$=0.076229825$\pm$0.000000025 days, with $P_{\rm post}$=0.07623366$\pm$0.00000060 days, for $\Delta P$=0.00000383$\pm$0.00000060 days.  The fractional period change is $\Delta P$/$P$ equals $+$50.3$\pm$7.9 parts-per-million (ppm).  Importantly, the kink does not happen near the start of the eruption (a time when the old model has the ejection being driven off by the initial thermonuclear blast), but rather the kink is at a time 108.4$\pm$12.6 days after the start of the eruption.  The steady period change (and it is very steady) before the eruption was $\dot{P}_{\rm pre}$=($+$6.49$\pm$0.07)$\times$10$^{-10}$ in dimensionless units, and after the eruption was $\dot{P}_{\rm post}$=($+$3.67$\pm$0.27)$\times$10$^{-10}$.
 
\section{MASS OF THE NOVA EJECTA}

The fate of T Pyx depends on whether $M_{\rm ejecta}$ is larger or smaller than $M_{\rm accreted}$ over the eruption cycle.  I have just worked out the complex history for $M_{\rm accreted}$ to useable accuracy.  Now we need to work out the complex history for $M_{\rm ejecta}$.  Unfortunately, all prior published claims to measure $M_{\rm ejecta}$ for novae have real errors of 2--3 orders-of-magnitude, and the particular values of T Pyx span a range of 130$\times$.  And the theory estimates of $M_{\rm ejecta}$ are also with real uncertainties at the 2--3 orders of magnitude level, and worse, they all have missed the dominant mechanism for the mass ejection, so all prior theory is irrelevant and misleading.  This means that to answer the science question, we need a new and reliable observational method to measure $M_{\rm ejecta}$.  The new method is to work from the measured $\Delta P$, with simple and reliable physics, to derive $M_{\rm ejecta}$.

\subsection{Prior claims for $M_{\rm ejecta}$}

All prior published observational measures of $M_{\rm ejecta}$ are based on the traditional method that use observed emission flux from hydrogen, converting this to a photon luminosity, then assuming some model for the shell to calculate an emission rate and the number of hydrogen atoms.  This general procedure has many variants, and can use the Balmer or Paschen emission or the radio flux.  This traditional method is the {\it only} method available for nearly all novae.  This traditional method is what we learned in graduate school, and it is now used so frequently that this has become the venerable and default model, used in most papers.  But universality and repetition does not mean that there is any useable accuracy in the traditional method.

For T Pyx in particular, the published estimates of $M_{\rm ejecta}$ vary widely:  Contini \& Prialnik (1997) used the H$\beta$ flux to estimate that T Pyx ejected 16$\times$10$^{-7}$ M$_{\odot}$.  Selvelli et al. (2008) claim that the T Pyx $M_{\rm ejecta}$ is (100--1000)$\times$10$^{-7}$ M$_{\odot}$.  Nelson et al. (2014) use a model of the radio emission to estimate the ejected mass to be (100--3000)$\times$10$^{-7}$ M$_{\odot}$.  Shore et al. (2013) used the H$\beta$ flux and the traditional method to estimate that the ejected mass of the 2011 eruption was $\approx$20$\times$10$^{-7}$ M$_{\odot}$.  Chomiuk et al. (2014) report that $M_{\rm ejecta}$ is $\gtrsim$100$\times$10$^{-7}$ M$_{\odot}$.  Pavana et al. (2019) use the traditional methods with a simplistic model to claim $M_{\rm ejecta}$=70.3$\times$10$^{-7}$ M$_{\odot}$, for remarkable precision and no error bars.  Izzo et al. (2024) use a complex model for the distribution of the ejecta along with the H$\beta$ flux to derive that the mass is $\ll$(30$\pm$10)$\times$10$^{-7}$ M$_{\odot}$, where the $\ll$ sign is for their statement that they did not include any correction for a filling factor and this is likely to lower the value by one order-of-magnitude.  These range over a factor of 190$\times$.  With this, we see the real uncertainty in the prior mass estimates, and we see that these attempts are useless for answering the question of whether $M_{\rm ejecta}$$>$$M_{\rm accreted}$.

For novae in general, the published $M_{\rm ejecta}$ values span huge ranges, even for estimates of the same nova eruption.  Here are two example novae, which have four or five published estimates, so we can see the full scope of the problem:  HR Del was an ordinary bright nova in 1967, with a 0.214 day period and a low mass WD.  Traditional observational methods produced estimates of 900 (Anderson \& Gallagher 1977), 2500 (Malakpur 1973), 15,000 (Robbins \& Sanyal 1978), and 1000--1500 (Tylenda 1979), all in units of 10$^{-7}$ M$_{\odot}$.  These span a range of 17$\times$.  U Sco is a recurrent nova with a high-mass WD and a subgiant companion star.  Traditional methods give published $M_{\rm ejecta}$ of 0.72 (Barlow et al. 1981), 0.1--1.0 (Williams et al. 1981), $\sim$1.0 (Anupama \& Dewangan 2000), 7.2--23 (Banerjee et al. 2010), and 21 (Pagnotta et al. 2015), all in units of 10$^{-7}$ M$_{\odot}$.  The range of these estimates is a factor of 230$\times$, which is to say that these traditional methods have accuracies that are not useful for any application.

All the traditional methods have many problems with large systematic errors.  These are listed in detail in Appendix A of Schaefer (2011):  {\bf First}, the recombination coefficient depends sensitively on the temperature of the expanding nova shell, with the temperatures known only approximately and it varies widely throughout the shell.  Williams et al. (1981) changed the assumed temperature over a reasonable range and the `measured' $M_{\rm ejecta}$ changed by a factor of 1500$\times$.  {\bf Second}, the filling factor has no evidential basis for which to measure it, so everyone always just puts in a guess, and these guesses vary by over a factor of 100$\times$.  {\bf Third}, the nova distance enters the the equations as a square factor, and in the old days the distances were always known with real uncertainties that when squared are a factor of 10$\times$ (Schaefer 2018).  Now with good {\it Gaia} parallaxes for the bright novae, this source of uncertainty is useably small, but all the pre-{\it Gaia} papers still have this large source of uncertainty.  {\bf Fourth}, the traditional methods explicitly assume that the shell is optically thin so that we see all the line flux, but the shell is optically thick up until the transition time long after peak, and the lines remain optically thick for much longer.  Usually, the papers apply the traditional method to times before the nova transition, so that all the photons from inside the photosphere are lost, all the volumes being shadowed by the photosphere are lost, and all the photons from the inside of the assumed clumps in the shell are lost, even at late times. This under-estimation of $M_{\rm ejecta}$ depends on details of the time of observations and the largely-unknown details of the mass distribution.  So in general, the traditional measures have systematic errors possibly up to many orders-of-magnitude.  {\bf Fifth}, the `measured' mass depends as the square of the electron density, for which typical constraints have 2--4 orders of magnitude uncertainty.  And the gas density and degree of ionization must vary greatly throughout the shell, so the assumption of any average electron density cannot accurately represent this highly non-linear situation.  So this fifth problem for the traditional methods has typical errors of $>$2 orders-of-magnitude squared.  {\bf Sixth}, the traditional methods all scale the `measured' $M_{\rm ejecta}$ with some adopted shell volume, but this is a problem because shell volumes are fuzzy-edged and holey-inside and hard to measure.  Most papers applying the traditional method simply take the volume as scaling with the cube of some radius derived from some velocity taken somehow from a line profile.  But line profile widths change by up to factors of two throughout the eruption, and it is not clear as to when the velocity should be measured to correspond to the volume.  Further, no paper ever justifies whether it is appropriate to use the HWHM or the HWZI for the usually-complex line profiles.  Further, the line-of-sight velocity (used to calculate the volume) depends on the unknown inclination with respect to the bipolar ejecta.  Further, real shells are toruses and bipolar outflows, and occupy only a fraction of the idealized spherical volume for the shell.  For the sixth problem, all these systematic errors must introduce uncertainty at the level of $\gg$$2^3$$\times$.  Most papers applying the traditional methods ignore most of these real systematic problems, and thus quote unrealistically-small error bars, while the majority of papers do not quote any error bars at all.  In the end, with these six major problems, the real uncertainty in measures of nova $M_{\rm ejecta}$ values are 2--3 orders of magnitude or more.

Theory estimates and modeling of $M_{\rm ejecta}$ also have 2--3 orders of magnitude in real uncertainty.  Kato, Hachisu, \& Saio (2017) chronicle the large effects on $M_{\rm ejecta}$ for many free choices of the modelers.  Critical free assumptions include the number of nova cycles followed by the model, whether helium flashes are used, the mass loss algorithm, and the number of layers into which the mass grid is divided.  For this last parameter, Starrfield et al. (2020, Appendix Table A1) quote their own study where they only changed the number of layers that the exploding layers were divided into as used in the calculation, from 95 to 150 to 200 to 300, with resulting $M_{\rm ejecta}$ values of 0.024, 0.36, 0.19, and 0.57 times $10^{-7}$ M$_{\odot}$.  This spans a range of 24$\times$, with no trend or sign of convergence.  And this is just one parameter out of many for which the computer programmer can freely choose.  This one result alone shows that theoretical models are not reliable with any usable accuracy.

Unfortunately, theory models have the calculated $M_{\rm ejecta}$ depending sensitively and critically on astrophysical parameters that are largely unknown.  Starrfield et al. (2025) present a study for one case (for a 1.35 $M_{\odot}$ ONe WD) where one parameter at a time is changed over the plausible range.  For a changing WD radius from 1522, 1827, and 2166 km, the ejected mass changes from 0.53, 1.07, and 0.12 (in units of 10$^{-7}$ M$_{\odot}$), for a factor 9$\times$ just for not having the exact WD radius for a given mass.  As the assumed oxygen fraction in the burning gas changes from 25\% to 50\%, $M_{\rm ejecta}$ changes from 0.15 to 1.02 times 10$^{-7}$ M$_{\odot}$, for a factor of 5$\times$ difference on an unknowable abundance.  And by changing the WD from CO to ONe, the ejecta changes from 0.53 to 4.2 times 10$^{-7}$ M$_{\odot}$, a factor of 8$\times$ for a fundamental property of the WD that can usually not be determined.  Starrfield et al. (2024) report on single-parameter variations for a different case (with a 0.60 $M_{\odot}$ CO WD), with more horrifying results.  Varying the WD mass within the typical one-sigma uncertainty for such CVs (see Schaefer 2025), the calculated $M_{\rm ejecta}$ changes by a factor of 4$\times$.  For unknowable changes in the mixing from dredge-up during the nova burning, $M_{\rm ejecta}$ changes from 4.9 to 1600 times 10$^{-7}$ M$_{\odot}$, a factor of 330$\times$.  The conclusion is that ordinary ranges of assumptions for unknowable astrophysics in the model inputs leads to real uncertainties of $>$300$\times$.  

In practice, we can get an estimate for the real uncertainty in theoretical model calculations for a single case by comparing published $M_{\rm ejecta}$ from independent groups of modelers.  Starrfield et al. (2020) report on four independent calculations for a 1.15 $M_{\odot}$ CO WD, with values of 150, 49, 0.098, and 130 times 10$^{-7}$ M$_{\odot}$, for a range of 1530$\times$.  For the case of U Sco, we find published ejecta masses of 2.1 (Kato 1990), $\sim$18 (Hachisu et al. 2000), 4.3 (Starrfield et al. 1988), 44 (Yaron et al. 2005), plus 11 and 21 (Figueira et al. 2025), covering a range of 21$\times$. 

So theory can only produce $M_{\rm ejecta}$ estimates with real total uncertainties from 100$\times$ to 1000$\times$.  But this is not the worst of the case against utility for theory in this one question.  The big over-riding problem is that all published theory estimates have completely missed the dominant mass ejection mechanism.  All prior models are made using the old 1-dimensional models where mass loss is from ballistic ejection by an optically thick wind.  Rather, we now know that the dominant mass ejection mechanism is from the dynamical effects of the companion star swirling around in the outer part of the hot envelope surrounding the white dwarf (Chomiuk, Metzger, \& Shen 2020, Sparks \& Sion 2021, Shen \& Quataert 2022).  By missing out on the dominant mass ejection mechanism, all prior theoretical models can only be relegated.

I am reciting these facts of woe, because most workers have no realization that the many confidently-published $M_{\rm ejecta}$ values all have real uncertainties of two-to-three orders-of-magnitude.    I am also detailing these condemning facts to justify the desperate need for some new method that is both accurate and reliable.

\subsection{Time Structure For Mass Ejection}

Schaefer et al. (2013) collect a well-resolved light curve of the entire rise from background for the 2011 eruption.  This includes the two earliest magnitudes ($B$=12.51 and $B$=12.50) above the quiescent level, taken fortuitously by myself as part of service observing with the 0.9-m SMARTS telescope on Cerro Tololo as scheduled several months in advance.  The first 2.0 days of the eruption have the many magnitudes exactly fitting a simple model with a linear expansion velocity of a constant-temperature photosphere.  That is, the nova flux follows a time-squared equation.  At Day +2.0 of the eruption, the parabola breaks into a nearly flat light curve, presumably now with a receding photosphere.  This is strong evidence of a large ejection of gas around Day-zero of the eruption.  This fits the olden schematic picture that nova ejecta are entirely blasted off the white dwarf in the first few minutes of the runaway thermonuclear explosion, with this impulsive ejection stopping due to the expansion of the accreted layer that has just exploded.

Schaefer (2023b) reports on 129 photometric minimum times for T Pyx from 1986--2022, including from my many time series mostly from Cerro Tololo, many times of minimum as reported in the AAVSO International Database, as well as with {\it TESS}.  The pre-eruption and post-eruption orbital periods and steady period changes are defined with extremely good accuracy.  The $O-C$ curve from Figure 7 of Schaefer (2023b) is extended in Figure 4 to illustrate the situation of the changing period and to demonstrate the high quality of the timings.  For this sub-section, the important point is to note that the best-fitting pre- and post-eruption $O-C$ parabolic curves cross at Day +108.4$\pm$12.6 days after the start of the eruption.  The slope of the $O-C$ curve gives the instantaneous orbital period, and the orbital period changes instantaneously as the ejected gas passed outside the orbit of the companion.  For an instantaneous ejection of gas, the $O-C$ curve will have a sharp kink on the day of the ejection.  For an ejection of gas over a range of days, the crossing of the pre- and post-eruption parabolas will be at the time of the halfway point in the ejection.  So the observed $O-C$ curve proves that the middle time of the gas ejection is on Day +108.4 days.  This could be because the majority of the gas ejection was close to Day +108.4, or possibly uniformly from Days +70 to +147, or possibly with exponentially-falling ejection from some early time in the eruption up to Day +300 and later.  This later ejection must be distinct from the initial blastwave, by some different physical mechanism.

Surina et al. (2014) present an excellent analysis of their comprehensive photometry and spectroscopy from the entire 2011 eruption, and they find distinct phases for the gas ejection.  They ``propose two different stages of mass loss, a short-lived phase occurring immediately after outburst and lasting $\sim$6 days, followed by a more steadily evolving and higher mass loss phase.''  The early rise phase had an initial ejection velocity of 4000 km/s, but this drops to $\sim$2000 km/s by Day +2.7.  The initial rise is seen to last up to Day +3.3, with the expansion described as a Hubble flow.  T Pyx next went into a `pre-maximum halt' phase from Days +3.6 to +13.7, with this being the start of a ``steadily evolving and higher mass loss phase''.  The `final rise' phase is from Day +14.7 to the peak on Day +27.9, featuring a gradual increase in the ejection velocity.  Such an increase is not possible with the initial blastwave, thus requiring some second mode of continuing ejection.  The `early decline' from Days +27.9 to +90 and the `transition to the nebular phase' from Days +90 to +280 chronicles the emergence of the Orion spectrum and its transition to the optically thin nebular spectrum.  The last three phases require the continual replenishment of ejecta, as a separate mode of ejection independent of the initial blastwave.

Nelson et al. (2014) model the radio emission flux at four frequencies from Days +8 to +528 of the 2011 eruption.  They were startled to find a low flux in the first 76 days.  They offer two possible explanation.  One possibilities is that the initial ejecta is `cold', but such seems unphysical.  The only realistic possibility is that ejection of gas was thin over the first 76 days, which is to say that most of the ejected gas comes out long after the initial thermonuclear runaway explosion.  They model this as all the ejecta being sent out on Day +76, and this well fits their radio light curves.  The radio light curves allows for some small fraction of the ejecta to come in early phases so as to explain the faint early flux.  Further, the radio light curves provide only modest constraints on continuing ejection after Day +76.

Chomiuk et al. (2014) use the X-ray light curve to argue for two distinct episodes of ejection.  The observational basis for this is that T Pyx shows a rising {\it hard} X-ray flux starting on Day +117, with the flux brightening to a maximum on Day +142, stays on a flat plateau until day $\sim$206, and then gradually fades.  They strongly argue that this hard flux can only arise from internal shocks within the ejecta.  They model this as two brief episodes of gas ejection, with the first on Day-zero with a velocity $\sim$1900 km/s and the second ejection around Day +75 with a velocity of 3000 km/s.  This model has the strong advantage of explaining the maximum blueshift in the H$\beta$ line of $\sim$1900 km/s on Day +2 and $\sim$3000 km/s on Day +69.  Further, this model explains the radio light curves.  The {\it model} has instantaneous ejection of the second shell on Day +75, but there inevitably must be a substantial spread in time of this second ejection.  This can be seen with the H$\beta$ velocity already being in the high-state on Day +69, and by the hard X-ray flux continuing brightly even past Day +300.

All of these data ($P$, optical, radio, and X-ray) are forming a unified picture of how the mass was ejected during the 2011 eruption.  T Pyx had two distinct modes of ejection.  The first mode is a blastwave from the explosion by the triggered thermonuclear runaway.  The second mode is a separate continuous ejection, starting sometime around the early days, peaking in ejection rate perhaps over Days +70 to +110, and fading off with ejections out to Day +300 and later.  The old paradigm for nova eruptions was entirely from the first mode.  The second mode is an addition to the first mode.  I think that T Pyx in 2011 was the first system in which the two-modes of ejection were recognized.

The physical mechanism of the first ejection mode was simply the traditional idea of runaway thermonuclear burning in the original layer of accreted gas making a blastwave.  Such runaways produce large amounts of energy on a timescale of hours.  The runaway ends when the accreted layer expands from the rapid heating to the point where the density drops sufficiently low (releasing the degeneracy of the gas) so as to limit the rate of burning.  This initial input of energy is what powers a massive ejection, driven outwards in the first few hours of the explosion.  The expanding shell has a fairly sharply defined photosphere, and it is the linear expansion of this photosphere that causes the rise in brightness from quiescence with a time-squared dependency.  In the phase named as the `pre-maximum halt', the photosphere starts receding into the explosive Hubble flow, stopping the fast initial rise in brightness, and resulting in a lowering of the expansion velocity recorded for the shell's photosphere.  Even though the runaway explosion lasts only for a few hours, its expanding photosphere will dominate the optical light for the first week

The physical mechanism for the second mode (the long-running continuous ejection) was first proposed by Chomiuk et al. (2014), with a detailed physics model in Shen \& Quataert (2022).  The mechanism operates with the hot envelope surrounding the white dwarf that was formed from the original accreted material expanding greatly upon being heated by the nova.  This same hot envelope is what will become visible as the supersoft X-ray source.  They start off by demonstrating that continuous ejection from a wind on the surface of the puffed-up hot envelope is impossible in most cases because the companion stars are so close that the wind-launch regions are cutoff.  Rather, the idea is that the close-in motion of the companion star within the envelope will make for binary-driven mass loss as the companion stirs the envelope from the inside.  The ordinary motion of the companion star as it moves through this expanded envelope will necessarily suffer gas drag that will rob angular momentum from the orbit.  The companion star will churn the outer gas in the envelope and eject substantial gases each orbit.  This ejection mechanism operates from the start of the eruption up until the collapse of the envelope at the end of the super-soft phase, and the long duration means that large amounts of gas will be ejected by this mechanism.  For the case of T Pyx, the supersoft phase continues out past Day +300, so this explains the late ejections resulting in a middle date for ejection of Day +108.4.  For the case of T Pyx with a period {\it below} the Period Gap ($P$=0.0762 days), the companion orbits particularly close to the white dwarf, so it is especially strong at stirring and ejecting the envelope gas, thus explaining the huge observed $M_{\rm ejecta}$.

What is the time distribution of ejection for the second mode?  Presumably, this will be a function of the gas density around the orbital path of the companion star.  At this time, such cannot be predicted by theory.  The primary constraint is that the middle time for the ejection is near Day +108.4.  The optical data analyzed by Surina et al. point to the second-mode ejections starting early in the eruption (well before the peak on Day +27.9).  The radio and X-ray light curves indicate that the mass ejection before Day +75 was relatively small compared to the later ejections.  Further, the strength of ejection apparently picks up sharply around Day +75.  After the expanding nova shell goes transparent, the brightness of the supersoft X-ray source is telling us about the size/density of the hot envelope, so is also an indication of the mass ejection rate.  With this, the peak of the second-mode ejection was sometime roughly around Day +100.  With this, T Pyx was still ejecting gas up until Day +360.  For second-mode ejections running continually from Days +10 to +360 (peaking somewhere perhaps around Day +100), the middle date of ejection can easily be the observed Day +108.4$\pm$ 12.6.

What are the relative gas masses ejected by the two separate modes?  The relative strengths have importance for understanding the contributions to the orbital period changes across the eruption.    From the timing of the kink in the $O-C$ curve, we know that the initial ejection must be substantially smaller than the later continuing ejection.  Perhaps detailed physical models of the optical/radio/X-ray data can quantify the relative masses ejected by the initial blastwave versus the long-continuous ejection.  Nevertheless, the conclusion that only a  small fraction of the mass is ejected in the initial blast points to $\Delta$P$_{\rm AML}$ being dominated by the mechanism involving the companion orbiting inside the hot envelope.  

\subsection{$M_{\rm ejecta}$ for the RN eruptions}

The primary science task of this paper is to measure $M_{\rm ejecta}$ for the 2011 eruption, as based on the measured $\Delta P$.  This method is dynamical, based on simple physics, with no dependency on distance, filling factors, extinction, or temperatures.  This method is a simple timing experiment, where the long time range (covering 187,493 orbits) provides exquisite accuracy.  

This method has three down-sides:  First, a vast effort was needed by dozens of observers and spacecrafts over 39 years to collect the necessary $O-C$ curve.  This down-side has now been overcome.  Second, given the difficulty in knowing the angular momentum loss during the eruption, the result will only be a {\it lower limit} on $M_{\rm ejecta}$.  Fortunately, this limit is adequate to definitively solve the issue.  Third, this method can only be useful for a small set of novae, those for which the period change can be measured, and those for which the period change is {\it positive}.  Fortunately, T Pyx has a measured $\Delta P$ and that is positive.  Fortunately, the {\it limit} is overkill for definitively solving the issue.  The resultant limit is strict, with the simple physics and observations allowing no chance for deviation.

The primary physics for this paper is the change in orbital period resulting from the simple mass ejection from the WD.  Thus,
\begin{equation}
\Delta P_{\rm ejecta} = 2  \frac{M_{\rm ejecta}}{M_{\rm comp}+M_{\rm WD}} P.
\end{equation}
This simple result has been derived by many workers, and it was old even in 1983 (Schaefer \& Patterson 1983).  This equation is derived only with Kepler's Law and conserving angular momentum.  There is no way to violate the equation.

The problem with this equation is that the observed period change ($\Delta P$) is different from the period change resulting from the ejection of mass ($\Delta P_{\rm ejecta}$).  The difference is the period change resulting from angular momentum loss (AML) by the binary during the eruption ($\Delta P_{\rm ejecta}$).  This AML is from both the frictional AML from the initial blast and the frictional AML caused by the companion star orbiting within the outer part of the hot envelope around the WD.  The observed period change comes only from these two mechanisms, 
\begin{equation}
\Delta P = \Delta P_{\rm ejecta} + \Delta P_{\rm AML}.
\end{equation}
For all binaries for all cases, 
\begin{eqnarray}
\Delta P_{\rm ejecta}>0, \nonumber \\
\Delta P_{\rm AML}<0.
\end{eqnarray}
The ejection of mass from the WD can only make the period {\it increase} (see Equation 2).  The AML effect can only {\it decrease} the period, because the binary cannot get any angular momentum from the outside.

With $\Delta P_{\rm ejecta}$ being always-positive and $\Delta P_{\rm AML}$ being always negative, the positive-or-negative sign of $\Delta P$ depends on a balance of the two mechanisms.  Both mechanisms should be proportional to the density of the hot envelope at the radius that the companion orbits.  That is, a doubling of the density makes for twice as much gas being ejected by the swirling caused by the companion's orbit, and a doubling of the density makes for twice as much angular momentum to be carried away by the gas.  The balance depends on the detailed physics, which theorists have not been able to solve yet.

Theory cannot come up with a value for $\Delta P_{\rm AML}$.  But it is adequate for the purposes of estimating $M_{\rm ejecta}$ to only use the strict limit that $\Delta P_{\rm AML}$$<$0.  Then with Equations 2, 3, and 4, we have 
\begin{equation}
M_{\rm ejecta} > 0.5(M_{\rm comp}+M_{\rm WD})\frac{\Delta P}{P}
\end{equation}
Now we can finally calculate a strict limit on the ejected mass from the 2011 eruption.  With the above period and masses, I find $M_{\rm ejecta}$$>$(354$\pm$55)$\times$10$^{-7}$ M$_{\odot}$.  Patterson, Oksanen, Kemp, et al. 2017) have measured a similar $\Delta P$, used Equation 5, and arrived at a similar $M_{\rm ejecta}$$\gtrsim$300$\times$10$^{-7}$ M$_{\odot}$.

While theory cannot give any reliable value for $\Delta P_{\rm AML}$, it can reveal that the magnitude must be large.  That is, $\Delta P_{\rm AML}$ must be greatly more negative than for most novae.  In particular, the $P$ is below the Period Gap, so the companion must be orbiting deep with the envelope, where the gas density is exponentially higher than for other novae of larger orbital sizes.  Further, T Pyx has the mass ejections lasting for up to 300 days, which is longer than expected for an `average' nova, so the AML should be larger than for most novae.  With the two effects being large, then $\Delta P_{\rm AML}$ should be large.  This means that $\Delta P_{\rm AML}\ll0$.  With this, the `$>$' sign in Equation 5 should be replaced by the `$\gg$' sign.  Then, we know that $M_{\rm ejecta}$$\gg$354$\times$10$^{-7}$ M$_{\odot}$.  

For the eruptions in $\sim$1883, 1890, 1902, 1920, 1944, and 1967, the $M_{\rm ejecta}$ should be similar each time.  These limits on $M_{\rm ejecta}$ are placed into Table 3, to allow for the accounting of the total mass ejected during the entire eruption cycle.

We can do better than this strict limit.  The improvement is that we have some real limits on $\Delta P_{\rm AML}$ for novae that have {\it negative} $\Delta P$.  For these novae, necessarily with $\Delta P_{\rm ejecta}$$>$0, we get a limit on the AML as $\Delta P_{\rm AML}$$<$$\Delta P$.  Recall that $\Delta P_{\rm AML}$ is necessarily negative, so these novae place {\it useful} limits on the AML.

For comparison that minimize the differences in the nova $P$, it is best to consider the fractional period change $\Delta P$/$P$, with convenient units of parts-per-million.  All the novae with negative-$\Delta P$ are my own measures, as summarized in Schaefer (2023b).  QZ Aur has $P$=0.358 days and the peak of the eruption was missed, but the pre-eruption and post-eruption $O-C$ is accurately measured with $\Delta P$/$P$ of $-$290.31$\pm$0.20 ppm, proving that $\Delta P_{\rm AML}$/$P$ is more negative than $-$290 ppm.  HR Del was a J(231) neon nova that produced a bright shell, with $P$=0.214 days, the measured $\Delta P$/$P$=$-$472.5$\pm$3.4 ppm, so $\Delta P_{\rm AML}$/$P$ is more negative than $-$472 ppm.  DQ Her is a prototype CN with a deep dust dip for D(100) in a neon nova with a bright shell and $P$=0.194 days, for which the extreme limit on $\Delta P_{\rm AML}$/$P$ is $-$4.45$\pm$0.03 ppm.  RR Pic was a J(122) nova with $P$=0.145 days and an expanding ejecta shell, with $\Delta P$/$P$=$-$2004.0$\pm$0.9 ppm, so $\Delta P_{\rm AML}$/$P$ is more negative than $-$2004 ppm.  V1017 Sgr had its eruption peak missed, and has a sub-giant companion star, with $\Delta P$/$P$=$-$268$\pm$48 ppm, so $\Delta P_{\rm AML}$/$P$$<$$-$268 ppm.  To put these numbers into one place, the limits on $\Delta P_{\rm AML}$/$P$ are $<$$-$290, $<$$-$472, $<$$-$4.5, $<$$-$2004, and $<$$-$268 ppm.  With the median from this sampling of ordinary novae, our best estimate is that T Pyx would have $\Delta P_{\rm AML}$/$P$ being more negative than $<$$-$290 ppm.  But the range is large, so we really cannot make any strong constraints.

Adopting $<$$-$290 ppm as our best estimate for $\Delta P_{\rm AML}$/$P$, we can apply Equation 3 to get a limit for T Pyx.  This gives $\Delta P_{\rm ejecta}$/$P$$>$340.3 ppm.  With Equation 5, we have the best estimate to be $>$2400$\times$10$^{-7}$ M$_{\odot}$, with large error bars.

In the end, for the T Pyx RN eruption of 2011, the best estimate is that $M_{\rm ejecta}$$>$2400 $\times$10$^{-7}$ M$_{\odot}$, while we have a strict and unbendable limit of $M_{\rm ejecta}$$\gg$354$\times$10$^{-7}$ M$_{\odot}$.

\subsection{$M_{\rm ejecta}$ for the $\sim$1866 classical nova eruption}

The existence and year of the {\it c.}1866 eruption is proven by watching with {\it HST} the expansion of the T Pyx shell, where the observed fractional expansion can be extrapolated back in time (Schaefer, Pagnotta, \& Shara 2010).  The knots in the shell were demonstrated to have no measurable deceleration.  This measure of the expansion age has no distance dependency.  With the distance to T Pyx, the expansion velocity is seen to be 500--715 km/s, with this being greatly smaller than is possible for any of the high-velocity RN ejecta.  This demonstrates that the 1866 event is an ordinary classical nova.  With this setting the scene for the entire RN-phase of the eruption cycle.

Further, Schaefer, Pagnotta, \& Shara (2010) estimated the mass of the 1866 ejecta to be $\sim$10$^{-4.5}$ M$_{\odot}$.  They say ``The flux in emission lines (like H$\alpha$) from all the knots should be proportional to the mass of the shell, or at least the ionized fraction of that mass. The standard method equates the shell mass to a product involving the total line flux, various atomic constants, the square of the distance, and average densities of hydrogen and free electrons (Gallagher \& Starrfield 1978). Unfortunately, this method is highly model dependent, requiring assumptions on the poorly known distance, the filling factor, the composition, and the presence of a steady state. Typical uncertainties in these assumptions lead to the order of magnitude errors.''  The particular calculation for the T Pyx shell involves a correction for the knots turning on and off (as observed) from the flash ionizations of the later RN eruptions, with this involving uncertain modeling.  From the discussion in Section 3.1, we know that such traditional methods to observationally measure $M_{\rm ejecta}$ have real total uncertainties of two-to-three orders of magnitude.  Taking all this to heart, the real total uncertainty is like $\pm$1.2 in the exponent.  So our measured $M_{\rm ejecta}$ for the 1866 eruption is 10$^{-4.5 \pm 1.2}$ M$_{\odot}$.  That is, the best estimate is 320 in units of 10$^{-7}$ M$_{\odot}$, while the maximum possible range is 20--5000 in the same units.  In Table 1, the extreme lower limit is represented as $\gg$20 in units of 10$^{-7}$ M$_{\odot}$.

\begin{table*}
	\tablenum{3}
	\centering
	\caption{$M_{\rm accreted}$ and $M_{\rm ejecta}$ for the full eruption cycle}
	\begin{tabular}{lrrrr}
		\hline
		Years 	&	$\dot{M}$  & $M_{\rm accreted}$ & Extreme $M_{\rm ejecta}$ & Best $M_{\rm ejecta}$ \\
		 	&	($10^{-7}$ $M_{\odot}$/year)  & ($10^{-7}$ $M_{\odot}$) & ($10^{-7}$ $M_{\odot}$) & ($10^{-7}$ $M_{\odot}$) \\
		\hline
{\it c.} 1866	&	...	&	...	&	$\gg$20	&	320	\\
$\sim$1866 to $\sim$1883	&	4.00	&	68.0	&	...	&	...	\\
{\it c.} 1883	&	...	&	...	&	0	&	$>$2400	\\
$\sim$1883 to 1890	&	1.80	&	12.6	&	...	&	...	\\
1890	&	...	&	...	&	$\gg$354	&	$>$2400	\\
1890 to 1902	&	1.07	&	14.1	&	...	&	...	\\
1902	&	...	&	...	&	$\gg$354	&	$>$2400	\\
1902 to 1920	&	0.90	&	16.2	&	...	&	...	\\
1920	&	...	&	...	&	$\gg$354	&	$>$2400	\\
1920 to 1944	&	0.78	&	19.2	&	...	&	...	\\
1944	&	...	&	...	&	$\gg$354	&	$>$2400	\\
1944 to 1967	&	0.92	&	19.6	&	...	&	...	\\
1967	&	...	&	...	&	$\gg$354	&	$>$2400	\\
1967 to 2011	&	0.50	&	22.2	&	...	&	...	\\
2011	&	...	&	...	&	$\gg$354	&	$>$2400	\\
2011 to 2025	&	0.34	&	3.0	&	...	&	...	\\
2026 to $\sim$2137	&	0.26$\rightarrow$0.003	&	6.5	&	...	&	...	\\
$\sim$2137 to $\sim$14,937	&	0.003	&	38.5	&	...	&	...	\\
	&		&	\underline{~~~~~~~~~~~~~~~~}	&	\underline{~~~~~~~~~~~~~~~~}	&	\underline{~~~~~~~~~~~~~~~~}	\\
	&	$M_{\rm accreted}$ =	&	220	&		&		\\
	&		&	$M_{\rm ejecta}$ =	&	$\gg$2144	&	$>$17120	\\
		\hline
	\end{tabular}
\end{table*}

\section{$M_{\rm accreted}$}

Section 2.8 has collected the best estimates of the gas mass accreted over one complete eruption cycle.  To be specific, this includes the accretion from the 1866 eruption, through the RN-phase to roughly the year 2137, and through the quiescent phase to the time (crudely around the year 14,937 AD) when the next classical nova eruption kick-starts a new eruption cycle. 

For purposes of estimating $M_{\rm accreted}$ over the entire eruption cycle, I will tabulate the accreted masses from the time of the 1866 classical nova eruption until the next classical nova eruption sometime around the year 14,937 AD.  These time intervals are listed in Table 3.  They sum up to 220$\times$10$^{-7}$ M$_{\odot}$.  So my best estimate for the entire eruption cycle of T Pyx is $M_{\rm accreted}$=220$\times$10$^{-7}$ M$_{\odot}$.

What is the uncertainty on this value of $M_{\rm accreted}$?  Well, the inter-eruption intervals from 1883 to 2011 have a 20\% error bar due to the calibration of the accretion rate.  There is a 4\% uncertainty caused by the Menzel Gap for the 1944--1967 interval.  The post-2011 $\dot{M}$ has large uncertainty, but the accreted mass has relatively small uncertainty because the value comes from the trigger mass that depends on the nova mass and well-known physics, with the uncertainty in the upward direction being only 17\%.  The uncertainty in the mass accreted from roughly 1866 to 1883 is hard to know, but more than a factor of two seems implausible.  Adding these uncertainties in quadrature, in the upward direction, the error bar is 71$\times$10$^{-7}$ M$_{\odot}$.  With moderate round to express the real uncertainty, the best estimate for $M_{\rm accreted}$ is 220$\times$10$^{-7}$ M$_{\odot}$, and it is not any larger than 290$\times$10$^{-7}$ M$_{\odot}$.

\section{COMPARING $M_{\rm ejecta}$ TO $M_{\rm accreted}$}

The previous analyses of T Pyx as a SNIa progenitor candidate have never looked beyond one of the RN eruptions.  But the accretion and ejection history of T Pyx is complex and highly-changing, so the old works have missed most of the scenario and are largely irrelevant now.  To see the whole picture, the primary data sources are the Harvard plate stacks 1890--1989 and the many long photometric light curves 1968--2025 by roughly two dozen amateur and professional observers.  This work and this Section are the culmination of my program started in 1983, with T Pyx itself being targeted starting 1987.  The question of whether $M_{\rm ejecta}$ is larger or smaller than $M_{\rm accreted}$ is central to the grand-challenge SNIa progenitor problem for one of the premier progenitor candidates.

Finally, we can do the comparison between $M_{\rm ejecta}$ and $M_{\rm accreted}$ over a complete eruption cycle.  Over an entire eruption cycle, T Pyx accretes 220 in units of 10$^{-7}$ M$_{\odot}$.  Any one of the classical or recurrent nova eruptions will be ejecting more mass than is accreted over the entire cycle.  Over one full eruption cycle, the best estimate for the total ejected mass is $>$17120 in units of 10$^{-7}$ M$_{\odot}$, while the smallest possible total ejected mass is $\gg$2144 in the same units.  So the best estimate is that $M_{\rm ejecta}$ is $>$77$\times$ the $M_{\rm accreted}$.  We see that $M_{\rm ejecta}$ $\gg$ 11.3$\times$$M_{\rm accreted}$, with this being the extreme utter minimum case.  There is no way around this strict limit.

\section{EVOLUTION OF T PYX}

\subsection{Evolution over one eruption cycle}

The schematic evolution of T Pyx over the current eruption cycle is clear.  We know that the whole RN-phase was kick-started by a classical nova eruption around the year 1866.  The $\dot{M}$ has been falling since before the 1890 RN eruption.  This makes inevitable that the current RN-phase will end in a century or so, as the accretion returns back to its normal quiescent state.  This quiescent state must have a very low accretion level, as appropriate for a $P$=0.0762 day orbit.  Once in the quiescent state, T Pyx will look like an unremarkable faint CV with a low accretion rate, one essentially indistinguishable from the myriad, not worthy of a second look.  Over many millennia, the WD will slowly accumulate more hydrogen rich gas, eventually building towards the trigger mass.  When the trigger mass is accumulated again, a classical nova erupts, and the cycle starts over again.

I can now put in the critical numbers for the properties of this eruption cycle.  The 1866 classical nova eruption ejected $10^{-4.5 \pm 1.2}$ $M_{\odot}$ of oxygen-rich ejecta at a velocity near 600 km/s.  This forms the shell around T Pyx, even though what we see now is the tattered remains after the fast RN ejecta blasts through making Rayleigh-Taylor instability knots.  The classical nova burned hot, and continuing nuclear burning was initiated, much like a very-long-duration supersoft phase.  This burning irradiates the companion's atmosphere, puffing it up, and driving high-$\dot{M}$.  At first, in the decades after 1866, the accretion was likely somewhat above the steady hydrogen burning threshold, so $\dot{M}$$\sim$4$\times$10$^{-7}$ M$_{\odot}$/year.  But the feedback loop (hydrogen-burning making high-accretion making hydrogen-burning...) is not perfectly efficient, so the hydrogen-burning rate in quiescence is fading away.  Round about the year 1876 or 1883, the accretion fell below the threshold for steady hydrogen burning, so unburnt hydrogen started accumulating on the surface of the WD.  After the first 7 years or so with $\dot{M}$ just under 2$\times$10$^{-7}$ M$_{\odot}$/year, enough hydrogen had accumulated to trigger the first RN eruption.  With the high-but-falling accretion rate, subsequent RN events were triggered, at longer and longer intervals, while the star brightness faded.  Just before the 2011 eruption, the $B$ magnitude was 15.40, while the accretion rate was (0.5$\pm$0.1)$\times$10$^{-7}$ M$_{\odot}$/year.  The 2011 eruption ejected $>$2400$\times$10$^{-7}$ M$_{\odot}$, and certainly $\gg$354$\times$10$^{-7}$ M$_{\odot}$.  Since 2011, T Pyx has been fading steadily at the rate of 0.044 mag/year.  The system will fade to its quiescent level of something like $B$=21.0 (with $\dot{M}$$\sim$0.003$\times$10$^{-7}$ M$_{\odot}$/year) around the year 2137.  Over the entire RN-phase, $P$ gets {\it longer} by 785 ppm.  After 2011, the T Pyx WD cannot accumulate the trigger mass until after a long duration in quiescence, so there is no real possibility of another eruption within millennia, with $\sim$12,800 quiet years.  The whole eruption cycle has a duration of order 13,000 years.  Over this entire cycle, the WD accretes 220$\times$10$^{-7}$ M$_{\odot}$.  Over the cycle, the best estimate is that the nova eruptions eject $>$17120$\times$10$^{-7}$ M$_{\odot}$, and there is an inviolable limit of $\gg$2144$\times$10$^{-7}$ M$_{\odot}$.

The RN-phase is unique\footnote{IM Nor is often reasonably associated with T Pyx, due to both being RNe with periods below the Period Gap, but evidence on IM Nor is sparse, so any connection is not conclusive.  In particular, we have no evidence for a prior classical nova event, frequent RN eruptions, a secular decline of the accretion rate, nor a normal quiescent phase.  IM Nor does not have an observed nova shell, nor a measured $\Delta P$.}.  The RN-phase is startling because T Pyx is under the Period Gap, so all experience says that the accretion rate should be amongst the lowest out of all CVs, the complete opposite of T Pyx having amongst the highest accretion rate.  During its quiescent phase, T Pyx would appear similar to the myriad of faint CVs scattered all over the sky.  So what makes T Pyx special, and what property makes for the RN-phase?  The answer can only be that T Pyx separates itself from the other CVs by having {\it both} a WD mass near the Chandrasekhar mass and an orbital period below the Period Gap.  Somehow, these two properties lead to a prolonged and vigorous supersoft phase of nuclear burning.  This basic scenario was advanced by Knigge, King, \& Patterson (2000), Schaefer \& Collazzi (2010), and Patterson et al. (2014).  Some such scenario must be the mechanism to power the RN-phase, but the details are not known and not modeled.  Schaefer \& Collazzi (2010) note that T Pyx and the V1500 Cyg class of novae usually have high magnetic fields on their WDs, so conclude that highly magnetized WDs are causally connected to the very-long-duration supersoft burning\footnote{T Pyx is one of a class of stars named `V1500 Cyg stars', for which nuclear burning lasts many decades after the eruption is long over.  This class is defined by having the post-eruption brightness decades after the eruption being up to 5 magnitudes brighter than the pre-eruption star.  This class is characterized by short orbital periods, long-lasting supersoft emission, and highly magnetic WDs.  In this class, not all of the members have all of these properties.  Quantitatively, T Pyx differs from the other V1500 Cyg stars in that its $M_{\rm WD}$ is by far the largest.  Still, the T Pyx RN-phase is wrapped up with the mechanism for the V1500 Cyg stars.}.  I speculate that the mechanism is that the magnetic field funnels the accreted gas onto a small areas on the surface of the WD, where the local accretion rate can be sufficiently high to allow for continuous steady hydrogen burning, with this nuclear energy then intensely irradiating the nearby companion so as to drive the high-$\dot{M}$ that keeps the steady hydrogen burning.  

\subsection{Past evolution of T Pyx}

In each cycle of $\sim$13,000 years, T Pyx increases its orbital period by 785 ppm during the RN-phase, the companion star loses 220$\times$10$^{-7}$ M$_{\odot}$ to accretion onto the WD, and the WD ejects something like 17120$\times$10$^{-7}$ M$_{\odot}$.   With these details of the current eruption cycle, we can backtrack the evolution of T Pyx.  Roughly 40 eruption cycles ago (half-a-million years or so), the WD mass was 1.40 $M_{\odot}$.  Likely, the WD did not start out with an initial mass that close to the Chandrasekhar mass,   So the age of the T Pyx system as a CV must be somewhat less than 40 cycles, or less than half-a-million years.

A key consequence of any attempt to backtrack the state of the T Pyx system is that the WD must have formed `recently' with a mass 1.33-to-1.40 $M_{\odot}$ or so.  With a formation mass of 1.33-to-1.40 $M_{\odot}$, the T Pyx WD must have formed as an ONe WD.  There is no way around this.

T Pyx has an ONe WD, but it is not a neon nova.  How can this be?  Simple, the eruption cycles started recently, so the current dredged-up mass is from the crust and upper-mantle of the WD, where there is little neon.  De Ger\'{o}nimo et al. (2019, Figure 1) calculate the internal chemical profiles for a 1.29 $M_{\odot}$ WD, with the neon abundance being low for the outer 3.2\% of the WD mass, while this outer material must be over one-third oxygen.  That is, the original T Pyx ONe WD had its outer 0.043 $M_{\odot}$ largely neon free.  For T Pyx ejecting of order 0.0017 $M_{\odot}$ over each eruption cycle, T Pyx will not be a neon nova for the first 25 or so eruption cycles (roughly a third-of-a-million years).  Rather, the dredged-up WD material will have a super-solar abundance of oxygen, exactly as observed by Chomiuk et al. (2014).  For a nominal $M_{\odot}$=1.33 $M_{\odot}$, and a formation within the last 25 cycles, the start-up mass of the newly made WD must be $<$1.37 $M_{\odot}$.

An interesting point with this backtracked evolution is that the binary must have come into contact with a short $P$ under the Period Gap.  That is, the eruption cycle must have started less than 25 cycles ago, a time when the period was near 0.0747 days.  So the $P$ of contact was between 0.0747 and 0.0762 days.  I am startled to have a CV come into contact at such a short period, because since the 1980s, the canonical CV evolutionary track for each individual CV starts out with periods near 0.25 days, evolves to pass through the Period Gap, down to the minimum period (near 0.053 days), then up as a period bouncer (Rappaport, Verbunt, \& Joss 1983 Figure 2, and Knigge et al. 2011 Figure 12).  Nevertheless, detailed models of the formation state for CVs allows for contact roughly uniformly from periods of 0.050 to 0.32 days (de Kool 1992, Figure 5).  So the short period at the time of contact is actually no problem.

Over the complete RN-phase, the T Pyx binary {\it increases} its period by 785 ppm.  Standard theory for CVs in quiescence below the Period Gap is that gravitational radiation is the only mechanism for AML, and this is negligible in size compared to the other period changes.  Despite this, Schaefer (2024, 2025b) finds that most CVs and XRBs below the Period Gap have their $\dot{P}$ dominated by some unknown mechanism that is negative for nearly half of the systems.  In this case, we have no useful information on the steady period change of T Pyx in the quiescent phase.  In any case, the precedents of all other CVs and XRBs has little likelihood of being relevant for the weird and unique case of T Pyx.  So the best idea is to adopt the gravitational radiation result, and acknowledge that the $\dot{P}$  in the quiescent phase might be substantially different.  With this, the total period change over the full eruption cycle is $\sim$785 ppm.

Over each complete eruption cycle, T Pyx has its period {\it increase}.  This runs contrary to the deeply held ideal that all CVs evolve from long-$P$ to short-$P$.  That is, since the 1980s, graduate students learn this pervasive truism, with the primary expositions being Rappaport, Verbunt, \& Joss (1983), Patterson (1984), and Knigge et al. (2011).  The general idea is that the AML loss by the binary (both in quiescence and in eruption) can only make for a shorter period, because there is no way to supply angular momentum from the outside.  I have never heard anyone even question the validity of this conventional wisdom.  But now we have T Pyx as a confident case where the long-term evolution is certainly running short-$P$ to long-$P$, from the time of first contact.  So T Pyx is a blatant counterexample to the commonplace idea.  Now, for the case of T Pyx, it is easy and reasonable to allow for it to be an exception, because T Pyx is so weird and unique that we can have little reason to think that the T Pyx experience is applicable to other CVs.

Nevertheless, the question is raised as to whether the long-to-short ideal is indeed universal.  To answer what fraction of CVs and XRBs have their evolution long-to-short, we need a sample of systems with $\Delta P$ and $\dot{P}$ measures.  One ready-made sample is the seven systems listed at the end of Section 1.  Of these, we do not have measures of period changes for the symbiotic stars or FQ Cir, so the useful sample size is only 5 novae.  {\bf V445 Pup} ejected an extremely massive shell, and its $P$ is {\it increasing} from beginning to end.  {\bf U Sco} has competing effects as the steady $\dot{P}$ works to decrease the period, while the $\Delta P$ works to increase the period, making for a complicated period change that depends critically on the chosen start and stop dates.  Rather, the long term average change is made clear with my 1945 minimum time, so the `smooth evolution' curve (Figure 6 of Schaefer 2025c) is best, showing that U Sco is {\it increasing} in $P$ over the long term.  {\bf T CrB} also has competing effects from $\dot{P}$ and $\Delta P$ of opposite signs.  From the best fit model, the periods in 1866 and 2025 are close, but formally the $P$ difference is $-$790 ppm.  So over two whole eruption cycles, T CrB has a {\it decreasing} period.  {\bf V1405 Pup} has its period {\it increasing} from 2013 to 2025.  {\bf T Pyx} has its period {\it increasing} over its long eruption cycle.  So out of my sample of 5 novae, 4 of them have the periods increasing.

Another sample is the 52 CVs in Schaefer (2024), from which I select out the 44 systems with main sequence companions.  In this sample, 22 have their period increasing across all the years of observation, while 22 have their periods decreasing.  These many systems span all CV classes, so we cannot claim special cases.  With 44 systems, the result has good statistical significance.   The median duration of the $O-C$ curves is 40 years, while the maximum is 145 years.  A determined advocate for the universality of the long-to-short evolution could imagine that the evolution on the millennial timescale is always long-to-short, while the decade-to-century of observations is just seeing jitter up-and-down that is superposed on the longterm evolution.  For half of the individual CVs, my $O-C$ curves catch the downward parabolas, while for the other half, my $O-C$ curves catch the upward parabolas.  Such a retroactive excuse is not impossible.  But such an explanation would be an admission that the evolution is being dominated by some unknown period-changing mechanism, such that CV evolution is really being driven by this newly-invented mechanism.  The new physics is invented solely to deny what is seen in the data.  It is a poor argument for an advocate to postulate some unknown dominating $\dot{P}$-changing mechanism to explain one new dataset, because such a postulate is just assuming the desired answer.  So, the old ideal of universal long-to-short evolution appears to be busted, while the reality in the sky is that half the CVs are undergoing short-to-long evolution.

\subsection{The future evolution of T Pyx}

The future of T Pyx can be extrapolated forward by one eruption cycle at a time.  In roughly 130 eruption cycles ($\sim$1.7 million years), the WD will be whittled down to 1.1 $M_{\odot}$ and there can be no RN-phase at any accretion rate.  Such a binary will just appear as a V1500 Cyg star (Schaefer \& Collazzi 2010), with a century-long fade to quiescence long after the eruption is finished.  After 500 cycles (near 6.5 million years), the WD will be whittled down to half-a-solar-mass, all with a period around 0.106 days for a companion mass near 0.15 $M_{\odot}$.  With the schematic evolution, the WD is completely evaporated after 800 cycles ($\sim$10 million years).  Somewhere between 500 and 800 cycles, the accretion and nova eruptions will stop, and we'll be left with T Pyx as a separated binary.  The fate of T Pyx is to evolve into a detached binary with a low-mass compact star and a low-mass red star, so as to be largely indistinguishable from a single brown dwarf.

\section{T PYX IS {\it NOT} A SUPERNOVA PROGENITOR}

With $M_{\rm ejecta}$ $\gg$ 11.3$\times$$M_{\rm accreted}$, we know that the WD mass is {\it decreasing} `fast'.  With the WD certainly evolving away from the Chandrasekhar mass, it is impossible for T Pyx to evolve into a Type Ia supernova.  With high confidence, we have proven that T Pyx is not a SNIa progenitor.

The T Pyx WD is currently at 1.33 $M_{\odot}$ and it is losing mass at an extremely high rate.  So looking back in time, the WD was even more massive, and it must have been formed at a mass just below the Chandrasekhar mass.  With this, the WD can only be an ONe WD, a young ONe WD.  The nova eruption will not be a neon nova because the mass eroded from the mantle of the WD does not have a neon composition, rather the dredged material will be primarily carbon and oxygen (de Geronimo et al. 2019).  With the WD core being of ONe composition, the T Pyx WD can never become a Type Ia supernova.


  

\vspace{5mm}
\facilities{AAVSO, DASCH, TESS}


{}

\end{document}